\documentclass[12pt]{article}
\usepackage{amsmath}
\usepackage{graphicx}
\usepackage{enumerate}
\usepackage{natbib}
\usepackage{url} 

\usepackage{chngcntr}

\usepackage[T1]{fontenc}
\usepackage[latin9]{inputenc}
\usepackage{subfigure}
\usepackage{multirow}
\usepackage{tikz}
\usepackage{amsmath}
\usepackage{amsthm}
\usepackage{amssymb}
\usepackage{amstext}
\usepackage{bm}
\usepackage{booktabs} 
\usepackage{color}
\usepackage{enumerate} 
\usepackage[shortlabels]{enumitem}
\usepackage{epsfig,epsf,psfrag}
\usepackage{epstopdf}
\usepackage{float}
\usepackage{graphics}
\usepackage{graphicx}
\usepackage{latexsym}
\usepackage{lscape}
\usepackage{mathabx} 
\usepackage{mathtools}
\usepackage{mathrsfs}
\usepackage{multirow}
\usepackage{natbib}
\usepackage{rotate}
\usepackage{setspace}
\usepackage{algorithm}
\usepackage{algorithmic}
\usepackage{sectsty}

\usepackage{xargs}[2008/03/08]
\usepackage{xcolor}

\usepackage{ifthen}
\usepackage[hidelinks]{hyperref}
\hypersetup{
	colorlinks,
	linkcolor={red!50!black},
	citecolor={blue!50!black},
	urlcolor={blue!80!black}
}

\sectionfont{\centering\bf\large}
\subsectionfont{\normalsize}

\usepackage{titlesec}

\newtheorem{prop}{Proposition}

\newtheorem{theorem}{Theorem}
\newtheorem{lemma}{Lemma}
\newtheorem{cor}{Corollary}
\newtheorem{remark}{Remark}

{
	\theoremstyle{remark}

	\newtheorem{assumption}{Assumption}
}

\newtheorem*{lem*}{Lemma}

\newcommand{\bea}{\begin{eqnarray*}}
\newcommand{\eea}{\end{eqnarray*}}
\newcommand{\be}{\begin{eqnarray}}
\newcommand{\ee}{\end{eqnarray}}
\newcommand{\beq}{\begin{equation}}
\newcommand{\eeq}{\end{equation}}

\newcommand{\bal}{\begin{equation}\aligned}
\newcommand{\eal}{\endaligned\end{equation}}
\newcommand{\bgt}{\begin{equation}\begin{gathered}}
\newcommand{\egt}{\end{gathered}\end{equation}}
\newcommand{\ed}{\end{document}}

\newcommand{\btab}{\begin{tabular}}
\newcommand{\etab}{\end{tabular}}

\newcommand{\bi}{\begin{itemize}}
\newcommand{\ei}{\end{itemize}}
\newcommand{\bfi}{\begin{figure}}
\newcommand{\efi}{\end{figure}}
\newcommand{\ben}{\begin{enumerate}}
\newcommand{\een}{\end{enumerate}}
\newcommand{\bay}{\begin{array}}
\newcommand{\eay}{\end{array}}

\newcommand{\F}{Fr\'echet }

\def\real{\mathbb{R}}
\def\bX{\textbf{X}}
\def\bx{\textbf{x}}
\def\bz{\textbf{z}}

\def\bco{\iffalse}

\newcommand{\no}{\noindent}
\newcommand{\bc}{\begin{center}}
\newcommand{\ec}{\end{center}}

\definecolor{DarkBlue}{rgb}{0,.08,.45}
\definecolor{DarkRed}{rgb}{.7,0,.4}
\definecolor{DarkGreen}{rgb}{0,0.4,0}

\def\red{\textcolor{red}}

\def\T{\top}

\newcommand{\bpmt}{\begin{pmatrix}}
	\newcommand{\epmt}{\end{pmatrix}}

\newcommand{\bsp}{\begin{split}}
	\newcommand{\esp}{\end{split}}

\newcommand{\bdes}{\begin{description}}
	\newcommand{\edes}{\end{description}}

\newcommand{\bass}{\begin{assumption}}
	\newcommand{\eass}{\end{assumption}}
\newcommand{\bthm}{\begin{thm}}
	\newcommand{\ethm}{\end{thm}}
\newcommand{\blem}{\begin{lem}}
	\newcommand{\elem}{\end{lem}}
\newcommand{\bcor}{\begin{cor}}
	\newcommand{\ecor}{\end{cor}}
\newcommand{\bpf}{\begin{proof}}
	\newcommand{\epf}{\end{proof}}

\def\i1{_{i1}}

\def\cov{{\rm cov}}

\def\var{{\rm var}}

\def\eps{\varepsilon}

\def \mbb {\mathbb}

\DeclareMathOperator*{\argmin}{argmin}

\def\real{\mbb{R}}

\def\o{\omega}
\def\O{\Omega}

\newcommand{\m}[1]{m_{\oplus}({#1})}
\newcommand{\hatm}[1]{\hat{m}_{\oplus}({#1})}
\newcommand{\om}{\omega_\oplus}
\newcommand{\hatom}{\hat{\omega}_\oplus}
\newcommand{\mz}[1]{m^{(0)}_{\oplus}({#1})}
\newcommand{\Rp}{R^2_{\oplus,2\mid 1}}
\newcommand{\hatmz}[1]{\hat{m}^{(0)}_{\oplus}({#1})}

\newcommand{\ltwoNorm}[2][]{%
	\ifthenelse{ \equal{#1}{} }
	{\ensuremath{\|#2\|}}
	{\ensuremath{\|#2\|_{#1}}}
} 

\newtheoremstyle{italicassumption}
  {3pt}          
  {3pt}          
  {}     
  {}             
  {\itshape}     
  {.}            
  {.5em}         
  {}             

\theoremstyle{italicassumption}
\newtheorem{assumptionprime}{Assumption}

\theoremstyle{plain} 
\usepackage{etoolbox}
\patchcmd{\assumption}{Assumption}{}{}{\errmessage{Could not remove assumption heading name}}
\patchcmd{\assumptionprime}{Assumption}{}{}{\errmessage{Could not remove assumptionprime heading name}}
\renewcommand{\theassumption}{(G\arabic{assumption})}

\usepackage{authblk}
\def\bco{\iffalse}
\newcommand{\blind}{1}

\begin{document}

	\def\spacingset#1{\renewcommand{\baselinestretch}%
		{#1}\small\normalsize} \spacingset{1}

%

\if1\blind
{
		

\title{\bf Inference for \F Regression}
\author{Wookyeong Song\textsuperscript{a}, Paromita Dubey\textsuperscript{b}, Hans-Georg M\"uller\textsuperscript{c}, and Alexander Petersen\textsuperscript{d}  \\
	\textsuperscript{a}Brown University \\
    \textsuperscript{b}University of Southern California \\
    \textsuperscript{c}University of California, Davis \\
    \textsuperscript{d}Brigham Young University }
    \date{}
\maketitle
} \fi

\if0\blind
{
\bigskip
\bigskip
\bigskip
\begin{center}
{\LARGE\bf Inference for \F Regression}
\end{center}
\medskip
} \fi

\bigskip
\begin{abstract}
Linear regression is  widely used to model relationships between responses and predictors. In modern applications, one encounters data where the responses are non-Euclidean random objects situated in a metric space, paired with Euclidean predictors.  Global \F regression generalizes linear regression to such general settings; however, statistical inference
has remained largely unexplored. We develop a significance test for the null hypothesis that the \F regression function does not depend on the predictors, addressing the challenge of an absence of linear operations in metric spaces. We also develop a test for the partial effect of a subset of the predictors in analogy to, but quite different from, the partial F-tests commonly used in classical linear regression under Gaussian assumptions. Key ideas are  to employ random multipliers to obtain non-degenerate null distributions for the proposed test statistics and the Cauchy combination method.
We obtain consistency and convergence results and demonstrate the finite sample performance of the proposed tests through simulations on network data represented by graph Laplacians and spherical data with geodesic distances. We further illustrate our method using transport networks arising from New York City taxi trip data and U.S. energy source compositional data.
\end{abstract}

\noindent%

\newpage
\spacingset{1.9} 

\section{Introduction}
        \label{chap1}

Recent developments in metric statistics and object data analysis have motivated the study of data residing in general metric spaces $(\Omega, d)$ \citep{dubey2024metric, marron2021object, tavakoli2019spatial, song2026non}. A frequently encountered setting in this context is modeling the regression relationship between a random object response $Y$ in $(\Omega, d)$ and a Euclidean predictor $\bX$ in $\real^{p}.$ Aiming at the conditional \F mean of the response given a fixed predictor $\bX =\bx,$ that is \begin{equation}
\m{\bx}= \argmin_{\o \in \Omega} E\left(d^2(\o,Y) \mid \bX = \bx\right),
\label{eq:freg}  
\end{equation} a framework known as \F regression has been developed in recent years \citep{mull:19:5}. In particular, for multivariate predictors,  global \F regression avoids the curse of dimensionality and provides an effective tool for interpolation and extrapolation of object-valued responses without requiring parameter tuning \citep{fan2024conditional}. 


Inference on the significance of predictor effects $\bX$ on the response $Y$  is a fundamental aspect of regression analysis.  While the classical linear regression model with Euclidean responses $Y$ remains one of the most widely used tools for data analysis and is supported by an extensive literature, significance testing in global \F regression, which generalizes linear regression to metric space-valued responses, remains largely unexplored, in spite of its practical importance. This includes both the overall test for significance and also partial testing for the effect of subsets of predictors or single predictors. 
The goal of this work is to develop an inferential framework for global \F regression. 

In general metric spaces, estimation theory for the global \F regression function has been studied in  \citet{mull:19:5,scho:19} and others, but corresponding inference methods remain limited. Recent studies have investigated inference for global \F regression for a few specific metric spaces with particular distances, including distribution-valued responses under the Wasserstein metric \citep{pete:21}, and covariance matrices under the Bures-Wasserstein metric, for the global  \citep{xu2025wasserstein} and partial  \citep{xu2025test} effects of covariates. Other related works address different inferential targets: \cite{van2025test} proposed a test for discontinuities of the conditional \F mean at a fixed point using local \F regression, \cite{kurisu2026empirical} extended likelihood-based inference for  \F means and local \F regression, but not for global \F regression, and \citet{bhattacharjee2023single} studied an inference method for the index direction in a single-index \F regression model. 

We consider here significance testing in general metric spaces. Such spaces include graph Laplacians that can be used to represent networks, distributional data that can be represented in Wasserstein or Fisher-Rao spaces, or  spherical data with geodesic distance that can be used to represent compositional data \citep{scea:14}.    We propose a test for the overall significance of the global \F regression model and also a  test for partial effects of additional covariates for a given \F regression model, generalizing the classical coefficient of determination $R^2$  to a \F $R$-squared as a measure of goodness-of-fit. To the best of our knowledge, this is the first study to develop overall and partial significance testing procedures for global \F regression in general metric spaces. 

 In Section \ref{sec:prelim}, we present notations and necessary concepts from \F regression. 
 We propose a test for global predictor effect in Section \ref{sec:global} and the partial test in Section \ref{sec:partial}. We illustrate the size and power performance of the proposed tests through simulation studies using network and spherical responses in Section \ref{sec:simul}. Finally we apply the proposed significance tests to New York City (NYC) taxi transport networks and U.S. electricity compositional data
 in Section \ref{sec:data}. All proofs and technical details are relegated to the Supplement.


\section{Preliminaries}
\label{sec:prelim}

Let $(\O,d)$ be a separable metric space. Consider a random object $Y$ taking values in $\O$ and a $p$-dimensional Euclidean predictor $\bX.$ Let $P$ denote the joint probability measure of the paired data $(\bX, Y)$ on $\real^p \times \Omega,$ and let $P_{\bX}$ and $P_Y$ denote the marginal distributions of $\bX$ and $Y,$ respectively. We assume that the conditional distributions $P_{\bX|Y}$ and $P_{Y|\bX}$ exist. 

Suppose we observe a sample of random objects $\{Y_1,Y_2,\dots,Y_n\}$ in $\O$. The well-known unconditional population \F mean $\o_\oplus$ \citep{frec:48} and its sample estimator $\hat{\o}_\oplus$ are 
\begin{equation} \label{eq: fmean}
\o_\oplus = \argmin_{\o \in \O} E\left(d^2(\o,Y)\right),  \ \  \hat{\o}_\oplus= \argmin_{\o \in \O} \sum_{i=1}^n d^2(\o,Y_i).
\end{equation}
Writing $\bm{\mu}=E(\bX)$ and $\Sigma=\var(\bX)$ for the mean vector and covariance matrix of the predictors,  
the global \F regression model \citep{mull:19:5} is
\begin{equation}
\m{\bx}= \argmin_{\o \in \O} M\left(\o,\bx\right),
\label{eq:glo_freg}
\end{equation}
where the \F regression objective function $M(\o,\bx)$ is defined as
\begin{equation}
\label{eq:criterion}
M(\o,\bx) = E \left(s(\bX,\bx) d^2(\o,Y) \right), \quad \ s(\textbf{z},\bx) = 1+(\textbf{z}-\bm{\mu})^\T \Sigma^{-1} (\bx-\bm{\mu}). 
\end{equation}
Here we assume that $\Sigma$ is invertible. The specific form of the weight function $s(\bX,\bx)$ is motivated by a generalization of the classical multiple linear regression model, in which the value minimizing the \F function corresponds to the conditional expectation of $Y$ given $\bX = \bx.$ That is, when $\O$ is $\real$, $d$ is the Euclidean metric and $E(Y\mid \bX=\bx) = \beta_0 + \bm{\beta}_1^\T(\bx-\bm{\mu})$, global \F regression reduces to the standard linear model:
\begin{equation}
     \argmin_{\o \in \real} E \left(s(\bX,\bx)  (\o-Y)^2 \right) = \beta_0 + \bm{\beta}^\T_1(\bx-\bm{\mu}).
     \label{eq:linear_reg_Euclidean}
\end{equation}
As the weight function satisfies $E_X(s(\bX,\bx))=1$ for all $\bx \in \real^p$, the global regression function $\m{\cdot}$ passes through $(\bm{\mu},\o_\oplus),$ as $s(\bX, \bm{\mu}) = 1.$ {We assume throughout that all \F means $\o_\oplus$ in \eqref{eq: fmean}  and the global \F regression $\m{\bx}$ in \eqref{eq:glo_freg} are unique, see \ref{ass: existence_and_uniqueness} and Remark \ref{rm: unique_Fmean} below.}

Suppose one observes paired data of the form $(\bX_1, Y_1), (\bX_2, Y_2), \dots, (\bX_n, Y_n) \sim P,$ which are independent and identically distributed (i.i.d.) replicates of the random pair $(\bX,Y)$ taking values in $\real^p \times \O$. The estimated global regression function $\hatm{\bx}$ is the $M$-estimator
\begin{equation}
\hatm{\bx}= \argmin_{\o \in \O} M_n(\o,\bx),
\label{eq:glo_freg_est}
\end{equation}
where $M_n(\o,\bx)$ is the sample estimate of the \F function $M(\o,\bx),$ 
\begin{equation}
\label{eq:criterion_est}
M_n(\o,\bx)= \frac{1}{n} \sum_{j=1}^{n} s_{j}(\bx) d^2(\o,Y_j),  \ \quad \ s_{j}(\bx)=1+(\bX_j-\bar{\bX})^{\mathrm{T}} \hat{\Sigma}^{-1} (\bx-\bar{\bX}). 
\end{equation}
Here $\bar{\bX}$ and $\hat{\Sigma}$ denote the sample mean vector and covariance matrix of $\{\bX_1, \bX_2, \dots, \bX_n\}$, respectively. Before developing inference concerning the significance of the global \F regression model, we provide several examples of metric spaces $(\O, d)$ that are covered by the proposed  framework. 
These include standard Euclidean spaces $\mathbb{R}^p$ and Hilbert spaces $\mathbb{H}$ of functional data situated in an $L^2$-space over a compact domain, and also  
phylogenetic tree spaces with the BHV metric \citep{billera2001geometry}. Additional examples are as follows.  

\no \textbf{Example 1} (Network data).
\textit{Consider the space of simple undirected networks with vertex set $V$ and edge weight set $E,$ where the edge weights are bounded, i.e., $0\le w_{ij}\le C,$ for some constant $C\ge0$ and for all $i, j = 1,\ldots,m.$ Each network in this space is uniquely represented by its graph Laplacian matrix $G = (g_{ij}),$ defined as $g_{ij} = -w_{ij}$ for $i \neq j,$ and $g_{ij} = \sum_{k \neq i} w_{ik},$ for $i = j.$ The collection of such Laplacian matrices forms the space \begin{equation}
    \mathbb{L}^{m \times m} = \{G \in \real^{m \times m} \mid G = G^{T},~ G\mathbf{1}_{m} = \mathbf{0}_{m},\quad -C \leq g_{ij} \leq 0 ~\text{ for } i \neq j\},
    \label{eq:net_space}
\end{equation}
which can be equipped with the Frobenius metric or the power metric \citep{zhou2022network}. } 

\no \textbf{Example 2} (Compositional data). \textit{Compositional data are located in the $(p-1)$-dimensional simplex, $\Delta^{p-1} = \{ (a_1, \ldots, a_{p}) \in \real^{p} \mid a_j \geq 0, ~j = 1,\ldots,p,~\sum_{i=1}^{p} a_{i} = 1\},$ as they consist of non-negative vectors whose components sum to one. Applying the component-wise square root transformation, $(a^{1/2}_1, \ldots, a^{1/2}_{p})$ maps the elements of this simplex to the positive orthant of the unit sphere, 
    $\mathbb{S}_{+}^{p-1} = \{(b_1, \ldots, b_p) \in \real^p \mid \sum_{j=1}^p b_j^2 = 1,~ b_j \geq 0 ~\text{ for all } j\},$
with the geodesic distance  $d(\textbf{b}_1 , \textbf{b}_2) = \arccos(\textbf{b}_1^{\textnormal{T}}\textbf{b}_2),$ for $\textbf{b}_1, \textbf{b}_2 \in \mathbb{S}^{p-1}_{+}$ \citep{scea:14}.} 

\no \textbf{Example 3} (Covariance matrices). \textit{The space of $q \times q$ symmetric positive definite matrices, $\mathcal{S}_{++}^{q} =  \{S \in \real^{q \times q} \mid S^\T = S, \ S \succ 0  \},$ forms a smooth manifold. This space can be equipped with a variety of metrics, including the Log-Euclidean metric \citep{arsi:07}, Cholesky metric \citep{dryden2009non}, Log-Cholesky metric \citep{lin2019riemannian} and Bures-Wasserstein metric \citep{xu2025wasserstein}.}

\no \textbf{Example 4} ($2$-Wasserstein space of one-dimensional distributions). \textit{Let $\mathcal{P}(\mathcal{D})$ denote the space of probability measures on a compact domain $\mathcal{D}$.
Define the space of probability measures with finite second moments as $\Omega_\mathcal{W} :=  \left\{ \nu \in \mathcal{P}(\mathcal{D}): \int_{\mathcal{D}} \lVert s \rVert^{2} d\nu(s) < \infty \right\}.$
In the case of univariate distributions, the $2$-Wasserstein distance has the explicit representation
     $ d(\nu_1,\nu_2) = ( \int_{0}^{1} \{F_{\nu_{1}}^{-1}(s) -  F_{\nu_{2}}^{-1}(s)\}^2 ds )^{1/2},$ where $F_{\nu_{1}}^{-1}(\cdot),$ $F_{\nu_{2}}^{-1}(\cdot)$ are the quantile functions of $\nu_{1}$ and $\nu_{2},$ respectively. The space of univariate probability distributions equipped with $2$-Wasserstein distance is both complete and separable.}

\section{Inference}\label{sec:method}

\subsection{Test of global regression effect}\label{sec:global}

Let $\bX$ be a $p$-dimensional predictor taking values in a compact and convex domain $\mathcal{D} \subset \real^p$. 
To test the overall significance of the global \F regression model $\m{\cdot}$ in \eqref{eq:glo_freg}, we consider the null hypothesis \begin{equation}
\label{h0:global}
H^{G}_{0} : \m{\bX} = \o_\oplus \quad \text{almost surely,}
\end{equation} 
i.e.,  that the regression function $\m{\cdot}$ is almost surely constant. This null hypothesis holds when the response $Y$ is independent of the predictor $\bX$. However, testing for independence is a stronger problem than testing for a \F regression effect, since constancy of the mean regression function does not imply independence between $\bX$ and $Y$. We illustrate this distinction through examples in the simulation studies in Section~\ref{sec:simul}.



{The proposed  global test is based on the \F function $M(\o,\bx)$ in \eqref{eq:criterion}. {Define 
\begin{align}
    D(P)
=
E\left(M(\omega_\oplus,\bX)\right)
-
E\left(M(m_\oplus(\bX),\bX)\right),
\label{eq:numerator_diff}
\end{align} where $P$ is the joint probability measure of $(\bX, Y)$. Since $m_\oplus(\bx)$ minimizes $M(\cdot,\bx)$, one has $D(P)\geq 0$, so that $D(P)$ can be thought of 
as the improvement in the \emph{regression} loss 
that is attained by including $\bX$ in the model, while  the \emph{prediction} loss is  $L(P)
=
E\left(d^2(Y,m_\oplus(\bX))\right)
\geq 0.$ We define the \textit{global \F $R$-squared} as 
\begin{equation}
R^2_\oplus= \frac{D(P)}{D(P) + L(P)},  \label{R2}
\end{equation} which  is well defined whenever $Y$ is not almost surely equal to $\o_{\oplus};$   if $D(P) + L(P) = 0,$ then $Y = \omega_\oplus$ almost surely. If $(\O, d)$ is a Euclidean space, the numerator $D(P)=E\left(d^2(\o_{\oplus},m_{\oplus}(\bX))\right)$ corresponds to the standard notion of variance of $Y$ explained by the regression $m_{\oplus}(\bX)$ and the denominator $D(P) + L(P) = E\left(d^2(\o_{\oplus}, Y)\right)$ to the total variance of $Y,$ thus in this special case 
 $R^2_\oplus$ \eqref{R2} reduces to the standard population $R^2$}; see Proposition~S.1 in Supplement Section S.2.1 for further details.}

\bco
By definition, for each fixed $\bx,$ the \F regression $\m{\bx}$ minimizes $M(\o,\bx)$ over all $\o \in \Omega$; that is $M(\m{\bx}, \bx) \leq M(\o, \bx),$ for any $\o \in \O.$ Under the null hypothesis $H^{G}_{0}$ in \eqref{h0:global}, however, it should hold that $M(\m{\bX},\bX) = M(\o_\oplus,\bX)$ almost surely.
To formalize the comparison between $M(\m{\bX},\bX)$ and $M(\o_\oplus,\bX)$, we define the \textit{global \F $R$-squared} as 
\begin{equation*}
R^2_\oplus=1-\frac{E\left(M(\m{\bX},\bX)\right)}{E\left(M(\om,\bX)\right)} = \frac{E\left(M(\om,\bX)\right)-E\left(M(\m{\bX},\bX)\right)}{E\left(M(\om,\bX)\right)}.
\end{equation*}
The quantity $R^2_\oplus$ can be viewed as a generalization of the classical population coefficient of determination. Specifically, when $\O = \mathbb{R}$ is equipped with the Euclidean metric $d,$ the $R^2_\oplus$ reduces to the standard population $R^2,$ which measures the proportion of variance in $Y$ explained by the regression $E(Y\mid \bX)$ (see Supplement Section S.2.1 for details): \begin{align*}
    R^2= 1- \frac{E(\var(Y\mid \bX))}{\var(Y)}= \frac{\var\left(E(Y\mid \bX)\right)}{ \var(Y)},
\end{align*} where the second equality follows from the law of total variance.
\fi


Under the null hypothesis $H^{G}_{0}$ in \eqref{h0:global}, that is $\mathbb{P} (\m{\bX} = \o_\oplus ) = 1,$ we have  $R^2_\oplus=0$. If $\mathbb{P} (\m{\bX} \neq \o_\oplus) > 0,$ then by definition there exists a set of positive probability on which $M(\m{\bX}, \bX) < M(\o_\oplus, \bX),$ implying that $0 < R^2_\oplus \le 1;$  the \F $R$-squared $R^2_\oplus$ in \eqref{R2} thus has the hallmarks of  a measure of goodness-of-fit for global \F regression. 

Estimation 
is straightforward by replacing population quantities with their empirical estimates. This leads to the sample version of the global \F $R$-squared,  
{
\begin{equation}
\hat{R}^2_\oplus= \frac{\sum_{i=1}^{n}M_n(\hatom, \bX_i)- \sum_{i=1}^{n}M_n(\hatm{\bX_i}, \bX_i)}{\sum_{i=1}^{n}M_n(\hatom, \bX_i)- \sum_{i=1}^{n}M_n(\hatm{\bX_i}, \bX_i) +\sum_{i=1}^{n}d^2(\hatm{\bX_i}, Y_i)},
\label{eq:FRsquared_est}
\end{equation}
}
where $M_n$ is the \F function estimate in \eqref{eq:criterion_est}. To arrive at a principled test, we require some  asymptotic properties of $\hat{R}^2_\oplus$. We 
introduce the following assumptions:

{
\begin{assumption}\label{ass: lipschitz_metric}
   The \F regression function $\m{\bx}$ in \eqref{eq:glo_freg} and its estimator $\hatm{\bx}$ in \eqref{eq:glo_freg_est} 
 exist for all $\bx \in \mathcal{D}$, the latter almost surely, and there exists some $\tilde{\bx} \in \mathcal{D}$ such that  $E\left(d^4(\m{\tilde{\bx}}, Y) \right) < \infty.$
\end{assumption}
}

{
\begin{assumption}\label{ass: total_boundedness}
    For all $r \in (0,1),$ there exists a totally bounded set $K_r \subset \Omega$ such that 
    $\liminf_{n \to \infty} P( \hatm{\bx} \in K_r \text{ for all } \bx \in\mathcal{D} ) \ge 1- r.$
\end{assumption}

\begin{remark}
      \ref{ass: total_boundedness} requires a totally bounded subset around the \F regression estimator $\hatm{\bx}$ with high probability, and  is actually substantially weaker than the corresponding assumption in  the original \F regression paper  \citep{mull:19:5}, where total boundedness  of the  metric space $(\Omega,d)$ was required.  This type of assumption is commonly used to establish consistency of  M-estimators using empirical process theory (for example, Theorem 3.2.2 of \cite{well:23}). A sufficient condition is that the $\epsilon$-balls $B^{\o}_{\epsilon} = \{ y \in \Omega : d(\o, y) \le \epsilon\}$ are totally bounded for every $\o \in \Omega$ and $\epsilon > 0,$ and for some fixed $\omega_0\in\Omega$, $\sup_{\bx\in\mathcal D}d\bigl(\hatm{\bx},\omega_0\bigr)=O_P(1).$ Thus, in finite-dimensional Euclidean spaces, it is sufficient to have $\sup_{\bx\in\mathcal D}\|\hatm{\bx}\|=O_P(1)$; in infinite-dimensional spaces, boundedness alone is insufficient in general and additional conditions may be needed for \ref{ass: total_boundedness} to be satisfied, depending on the specifics of the space $(\O,d)$.
      \end{remark}}

\begin{assumption} \label{ass: existence_and_uniqueness}
For any $\epsilon > 0,$ $\Delta(\eps)=\inf_{\bx \in \mathcal{D}} \inf_{d(\o, \m{\bx}) > \epsilon} M(\o,\bx)-M(\m{\bx},\bx) >0.$
\bco
\red{Check if we even need the following part.}
	There exists $\xi=\xi(\epsilon) > 0$ such that as $n \rightarrow \infty$
	\begin{equation*}
	P \left(\inf_{\lVert x \rVert \leq  \mathcal{B}} \inf_{d(\o,\hatm{\bx}) > \epsilon} M_n(\o,\bx)-M_n(\hatm{\bx},x) \geq \xi  \right) \rightarrow 1.
\end{equation*}
\fi
\end{assumption}

\ref{ass: existence_and_uniqueness} guarantees that the \F regression objective function $M(\omega,\bx)$ in \eqref{eq:criterion} is well separated near the minimum $M(\m{\bx},\bx)$ for each $\bx \in \mathcal{D}$ and thus implies the uniqueness of the \F regression function $\m{\bx}$ for each $\bx \in \mathcal{D}$. 
\ref{ass: lipschitz_metric}--\ref{ass: existence_and_uniqueness} imply that the estimator $\hatm{\bx}$ is uniformly consistent for $\m{\bx}$, that is, 
$\sup_{\bx \in \mathcal{D}} d(\hatm{\bx},\m{\bx})=o_P(1)$;
for details, see Lemma S.1  
in  Supplement S.3.


For functions $h_1, h_2:\, \mathbb{R}^{p} \times \Omega \to \mathbb{R},$ define 
    $\langle h_1, h_2 \rangle_{L_2(P)} = \int h_1(\textbf{z}, y)h_2(\textbf{z},y) dP(\textbf{z},y),$ and $\| h \|_{L_2(P)} =\langle h, h \rangle_{L_2(P)}^{1/2}.$
For any $\varepsilon > 0$ and any function class $\mathcal{G}$, the $L_2(P)$-bracketing number of $\mathcal{G}$ \citep{well:23}  is
\begin{align*}
    & N_{[]}(\varepsilon, \mathcal{G},L_2(P)) \\ & = \min{\left\lbrace B: [g^l_j,g^u_j],~ j=1, \dots, B, \ \text{such that} \ \| g^u_j-g^l_j\|_{L_2(P)} < \varepsilon \ \text{and} \ \mathcal{G} \subset \bigcup_{j=1}^B  [g^l_j,g^u_j] \right\rbrace},
\end{align*}
where a bracket $[g^l,g^u]$ denotes the set of all functions $g$ such that $g^l \leq g \leq g^u.$
We consider functions $g_{\bx,\o}: \mathbb{R}^{p} \times \O \rightarrow \mathbb{R}$ of the form $g_{\bx,\o}(\bz,y)= \{1+ (\bz-\bm{\mu})^\top \Sigma^{-1} (\bx-\bm{\mu})\} \left \lbrace d^2(\o,y) - d^2(\m{\bx}, y) \right\rbrace,$
and the associated function classes $\mathcal{G}_\delta=\{g_{\bx,\o}: (\bx,\o) \in U_{\delta}\},$ where $U_{\delta}= \{(\bx,\o): \bx \in \mathcal{D},~ \o \in B_\delta^{\m{\bx}}\}$ for $\delta > 0$. 

{\begin{assumption} \label{ass: entropy}
 For any constant $c>0$, the bracketing entropy integral \newline  
     ${J}(\delta)=\int_{0}^{1} \left(1+\log N_{[]}(c \epsilon \delta , \mathcal{G}_\delta, L_2(P))\right)^{1/2} d\epsilon$
 satisfies ${J}(\delta) = O(\delta^{-\eta})$ for some $ 0 < \eta < 1$. 
\end{assumption}}

\ref{ass: entropy} is motivated by empirical process theory and is required to control the tail behavior of the empirical process $M_n-M$ in a neighborhood of the minimizer of $M$ \citep{well:23}.
One can also  consider an entropy condition,  \[
J^\star(\delta) := \int_{0}^{1} \sup_{\bx \in \mathcal{D}} \left(\log N(c \epsilon\delta,B_{2\delta}^{\m{\bx}},d)\right)^{1/2} d\epsilon= O(\delta^{-\eta}),
\]
where 
$N(\varepsilon,\Omega',d)$ denotes the covering number of $\Omega'$ by balls of radius $\varepsilon$ for any $\varepsilon>0$ and $\Omega'\subseteq \Omega$.

\begin{prop}\label{prop: bracketing_entropy_integral_sufficient_condition} 
Suppose that $\bx \mapsto \m{\bx}$ is  H\"older-continuous locally uniformly in $\bx \in \mathcal{D}$, that is, there exist constants $0 < \beta \leq 1$, $C > 0$ and  $\epsilon > 0$ such that $d(\m{\bx_1},\m{\bx_2}) \leq C \|\bx_1-\bx_2\|^\beta$
whenever $\|\bx_1 -\bx_2\| \leq \epsilon$. Then under \ref{ass: lipschitz_metric}, 
$J^\star(\delta) = O(\delta^{-\eta})$ implies $J(\delta) = O(\delta^{-\eta}),$ for any $\eta > 0.$  
\end{prop}

Proposition~\ref{prop: bracketing_entropy_integral_sufficient_condition} implies that the condition $J^\star(\delta)=O(\delta^{-\eta})$ is sufficient for \ref{ass: entropy} to be satisfied.  This condition is often  easier to verify than \ref{ass: entropy}, since it imposes a condition directly on the metric space $\Omega$, rather than on the function class governing the empirical process $M_n - M$ as in \ref{ass: entropy}.

\begin{remark}
    \label{rm: unique_Fmean}
    \ref{ass: lipschitz_metric} with $\bx = \bm{\mu}$ and \ref{ass: existence_and_uniqueness} implies that the population \F mean $\om$ exists, and for any $\epsilon > 0,$ $\Delta'(\eps)=  \inf_{d(\o,\om) > \epsilon} \{E(d^2(\o,Y))-E(d^2(\om, Y))\} >0.$  
Under \ref{ass: lipschitz_metric}--\ref{ass: existence_and_uniqueness}, $\hatom$ is a consistent estimator of the population \F mean. If additionally  \ref{ass: entropy} holds, then $\frac{1}{n} \sum_{j=1}^n \{d^2(\hatom,Y_j) - d^2(\om,Y_j)\} = o_P(n^{-1/2})$; see  Lemma S.1 in Supplement Section S.3.
\end{remark}



\begin{lemma}	\label{lem:degeneracy_glo} Under \ref{ass: lipschitz_metric}--\ref{ass: entropy},
	\begin{equation*}
	\frac{1}{n} \sum_{i=1}^{n} M_n(\hatm{\bX_i},\bX_i)-\frac{1}{n} \sum_{i=1}^{n} M_n(\m{\bX_i},\bX_i)=o_P(n^{-1/2}).
	\end{equation*}
\end{lemma} 

{ Lemma \ref{lem:degeneracy_glo} implies that $n^{1/2}\hat{R}^2_\oplus = o_P(1)$ under $H^G_0$ in \eqref{h0:global}. This does not, by itself, preclude a nondegenerate limit under a faster scaling $n^{a},$ for $a > 1/2$. For example, in the Euclidean linear case, vector-space operations are available, and a quadratic expansion of the least-squares criterion yields a nondegenerate limit of the scaled $R$-squared $n\hat{R}^2,$ and similar arguments may apply for Hilbert space-valued responses \citep{shen2004f} or the space of univariate distributions equipped with the $2$-Wasserstein distance \citep{pete:21}. However for general metric spaces $(\Omega,d)$, which may be positively or negatively curved, such  scalings may not be generally available; 
see Supplement Section S.2.2 for further discussion.
}

{To construct a unified test that applies to a broader class of metric spaces $(\O,d)$, we introduce random multipliers that generate a nondegenerate first-order perturbation with a consistently estimable conditional variance, rather than relying on a scaling of $\hat{R}^2_{\oplus}$ with nondegenerate limit that is not available across general metric spaces.} Let $\phi_1, \ldots, \phi_n$ and $\psi_1, \ldots \psi_n$ be i.i.d. copies of a random variable $Z$ with mean $1$ and variance $1/2,$ and assume that the $\phi_i$ and the $\psi_i$ are independent of the data. In practice, one may assume that $Z$ has a Gaussian distribution; but this is optional.  Motivated by the independent multiplier limit theorem \citep{gine1990bootstrapping}, and applying it to the numerator of the sample \F $R$-squared $\hat{R}^2_\oplus$ in \eqref{eq:FRsquared_est}, we define the perturbed test statistic:
\begin{equation}
T^G_n= \left \lbrace \frac{1}{n} \sum_{i=1}^{n} \phi_i M_n(\hatom,\bX_i) -\frac{1}{n} \sum_{i=1}^{n} \psi_i M_n(\hatm{\bX_i},\bX_i) \right \rbrace.
\label{eq:test_stat_glo}
\end{equation} We assume that $\sigma_G^2=  \frac{1}{2}E\{M(\om, \bX)^2 + M(\m{\bX},\bX)^2 \} > 0.$
The following Lemma characterizes the distribution of $T^G_n$ under $H^G_0$ in \eqref{h0:global}.
\begin{lemma}
	\label{thm:null_glo} 	Under $H^G_0$ in \eqref{h0:global}, and \ref{ass: lipschitz_metric}--\ref{ass: entropy},   
	\begin{equation*}
	 n^{1/2}\frac{T^G_n}{\hat{\sigma}_G} \rightarrow N(0,1) \quad \text{in distribution}, 
	\end{equation*}
    where $\hat{\sigma}_G^2=  \frac{1}{2n} \sum_{i=1}^{n}  \{M_n(\hatom,\bX_i)^2 + M_n(\hatm{\bX_i},\bX_i)^2 \}$ is a consistent estimator of $\sigma_G^2.$
\end{lemma}

The above limit  distribution of the global \F test statistic $T^G_{n}$ is a consequence of employing independent random multipliers. 
One can then  generate $K$ randomized statistics $(T_{n,1}^G,\ldots,T_{n,K}^G)$ from the same sample using independent multiplier draws. Let $\tilde T_{n,k}^G=n^{1/2}T_{n,k}^G/\hat\sigma_G$ denote the corresponding $z$-scores, for $k=1,\ldots,K$. To aggregate evidence across these randomizations, we further adopt the Cauchy combination test idea \citep{liu2020cauchy} to their individual $p$-values, namely forming a weighted sum of Cauchy transformed individual $p$-values from repeated randomizations to obtain an overall $p$-value. { The  benefit of combining repeated randomizations is improved power for fixed alternatives by aggregating evidence while preserving asymptotic validity under the null; see Supplement Section S.2.3 for further discussion on  the effect of $K$ in theory and numerical implementations.}
For a fixed integer $K$, the Cauchy combination test statistic is 
\begin{equation}
     C^{G}_n = \sum_{k=1}^{K} c_{k} \text{tan}\{(2\Phi(\lvert \tilde{T}^{G}_{n,k} \rvert) - 3/2)\pi\},
    \label{eq:test_stat_Cauchy}
\end{equation}
where $\Phi$ is the standard normal cumulative distribution function (CDF), $c_{k}\ge 0$ and $\sum_{k=1}^{K} c_{k} = 1.$ In the following, \textit{Cauchy}$(0,1)$ stands for the standard Cauchy distribution.  
\begin{theorem}
    \label{thm:null_glo_cauchy}
    Under the assumptions in Lemma~\ref{thm:null_glo}, for the test statistic $C^{G}_n$ in \eqref{eq:test_stat_Cauchy}, as $n \rightarrow \infty$, $C^{G}_n \rightarrow \text{Cauchy}(0,1)$ in distribution.
\end{theorem}
\noindent The rejection region for testing $H^G_0$ in \eqref{h0:global} is { $\mathcal{R}_{n, G}= \left \lbrace C^{G}_n   > \rho_{\alpha} \right \rbrace$} with significance level $\alpha,$ and here $\rho_{\zeta}$ is the upper $\zeta$-quantile of \textit{Cauchy}$(0,1).$

To investigate the power of the test, we construct a sequence of {local} alternatives,
\begin{align}
    H^{G}_{1,n} = \{Q : D(Q) \ge a_n  \},
    \label{eq:contig_alt_glo}
\end{align} with $D(\cdot)$ as in \eqref{eq:numerator_diff}. 
We assume that $\Sigma=\operatorname{var}(\bX)$  is uniformly positive definite for all $n$ and $Q \in H^{G}_{1,n},$ and  define the corresponding power function as $\beta^G_n = Q \left( \mathcal{R}_{n,G} \right).$ The asymptotic behavior of the proposed test under $H^{G}_{1,n}$ is established in the following result. 

\begin{theorem} \label{thm: power_global}
Assume \ref{ass: lipschitz_metric}--\ref{ass: entropy} hold uniformly in $Q \in H^{G}_{1,n}$ in \eqref{eq:contig_alt_glo}, and let $a_n$ be a sequence such that $a_n \to 0$ and $n^{1/2}a_n \to \infty.$ Then, one has power $\inf_{Q \in H^G_{1,n}} \beta^G_n(Q) \to 1.$
\end{theorem}

Theorem \ref{thm: power_global} illustrates the asymptotic power of the proposed test under local alternatives, and shows the consistency under $H^{G}_{1,n}$ with $n^{1/2}a_n \to \infty$. { This result indicates that the broad applicability of the  proposed test across a large class of  metric space-valued responses  may entail a loss of local efficiency in comparison to  specific metric spaces with stronger structural assumptions.  For the  Euclidean linear regression model  \eqref{eq:linear_reg_Euclidean} in  finite-dimensional Euclidean spaces, $D(P) = \bm{\beta}^\T_1 \Sigma \bm{\beta}_1,$ and consequently this condition corresponds to $\bm{\beta}_1$ converging to $0$ more slowly than
$n^{-1/4},$ under a nonsingular design covariance, whereas the classical F-test can detect coefficients of order $n^{-1/2}.$ Thus, the proposed test trades local efficiency for applicability to a broad class of response metric spaces $(\O, d)$.
}

\subsection{Test of partial effect}\label{sec:partial}

A common task in regression analysis  
is to assess whether incorporating additional predictors leads to a meaningful improvement in model fit; in linear regression a typical approach is to utilize F-tests under Gaussianity.  We extend this approach to 
global \F regression.  
Let   $\bX_1 \in \real^{p_1}$ be the predictors in a base model and $\bX_2\in \real^{p_2}$ the additional predictors to be considered.  The full predictor vector is  $\bX=(\bX_1^{\T},\bX_2^{\T})^{\T}$, and we assume that $\bX$ takes values in a  compact and convex set $\mathcal{D} = \mathcal{D}_1 \times \mathcal{D}_2 \subset \real^{p_1}\times \real^{p_2},$ with mean vector $\bm{\mu} = (\bm{\mu}_1^{\T}, \bm{\mu}_2^{\T})^{\T}$ and covariance matrix $\Sigma = (\Sigma_{kl}),$ where $\Sigma_{kl} = \cov(\bX_k, \bX_l),$ $k,l = 1, 2.$ 
 
We first consider the base \F regression model that utilizes predictors  $\bX_1.$ 
For fixed $\bx_1,$ define the corresponding objective function 
$M^{(0)}(\o,\bx_1)$ and \F regression $\mz{\bx_1}$,
\begin{equation}
\label{eq:base_criterion}
\mz{\bx_1}=\argmin_{\o \in \O} M^{(0)}(\o,\bx_1), \quad M^{(0)}(\o,\bx_1)= E \left(s^{(0)}(\bX_1,{\bx_1})d^2(\o,Y)\right),  \ 
\end{equation} with weight  function $s^{(0)}(\bX_1,{\bx_1})=1+({\bX_1}-{\bm{\mu}_1})^\top {\Sigma}_{11}^{-1} (\bx_1-\bm{\mu}_1).$


To test the partial effect of the additional predictors $\bX_2$ in the global \F regression, we consider the following null hypothesis,
\begin{align}
    \label{h0:partial}
H^P_0: \m{\bX}=\mz{\bX_1} \quad \text{almost surely,}
\end{align}
where $\m{\bX}$ in \eqref{eq:glo_freg} is the \F regression function under the full model using $\bX = (\bX_1^\T, \bX_2^{\T})^{\T}$, and $\mz{\bX_1}$ in \eqref{eq:base_criterion} under the reduced model using only $\bX_1.$ 

 In analogy to  the global test, { we define the partial improvement in the \F regression objective as
\begin{align}
D_{2\mid 1}(P)
=
E\left(M(m_\oplus^{(0)}(\bX_1),\bX)\right)
-
E\left(M(m_\oplus(\bX),\bX)\right).
    \label{eq:numerator_diff_partial}
\end{align} Pointwise minimization of $M(\cdot,x)$ implies that
$D_{2\mid 1}(P)\geq 0$. Define \textit{partial Fr\'echet $R$-squared} as
\begin{equation}
R_{\oplus,\,2\mid 1}^2
=
\frac{D_{2\mid 1}(P)}
{D_{2\mid 1}(P)+L(P)},  \label{Rpart} 
\end{equation} 
which  is well defined as long as $Y$ is not almost surely equal to $m_\oplus^{(0)}(\bX_1)$, as $D_{2\mid 1}(P) + L(P) = 0$ implies $Y = m_\oplus^{(0)}(\bX_1)$ almost surely. When $(\Omega, d)$ is Euclidean,  $\Rp$ \eqref{Rpart} boils down to the usual population partial $R^2$ for adding $X_2$ to a model containing $X_1$; see Proposition S.1 in Supplement Section S.2.1 for details.}

Under the null hypothesis $H^P_0$ in \eqref{h0:partial}, that is, $\mathbb{P}(\m{\bX} = \mz{\bX_1}) =1,$ we have $D_{2\mid1}(P) = 0$ and hence $\Rp=0$. If $\mathbb{P}(\m{\bX} \neq \mz{\bX_1}) > 0,$ then by definition $M(\m{\bX}, \bX) < M(\mz{\bX_1}, \bX)$ on a set of positive probability, which implies $0 < \Rp \le 1.$ In this case, the additional predictors $\bX_2$ are relevant for the \F regression.
\bco
{\color{blue} Counter example: 
$E\{M(\mz{\bX_1},\bX)\} < 0?$
Let $\bX_1 = (U,V)$ and $\bX_2 = W,$ with $(U,V,W) \sim \text{Bernoulli}(1/2).$
Let $\Omega = \{\o_{00}, \o_{10}, \o_{01}, \o_{11} \}$ with metric $d(\o_{uv}, \o_{u'v'}) = \lvert u -v\rvert + \lvert u'- v'\rvert.$ Define $Y = \o_{UV}.$ Consider $M_{1}(\o, \bx) =E(s(\bX, \bx)d^2(\o, Y)).$ Here, $\mz{\bX_1} = \o_{UV}$ and $M(\mz{\bX_1}, \bX)= -\frac{1}{2}$ almost surely, so $E(M(\mz{\bX_1}, \bX)) = -\frac{1}{2} < 0.$
\begin{align*}
    E_{(X,Y)}\left[M(\mz{\bX_1}, \bX)\right] &= E_{(X,Y)} \left[ E_{(X',Y')}\left\{s(\bX', \bX)d^2(\mz{\bX_1}, Y')\right\} \right] \\
    &= E_{(X,Y)} \left[ E_{(X',Y')}\left\{s(\bX', \bX)d^2(\o_{UV}, \o_{U'V'})\right\} \right]
\end{align*}
}
\fi
	To estimate the base \F regression $\mz{\cdot}$ under the null hypothesis $H^P_0$ in \eqref{h0:partial}, we use only the base predictors. Let $(\bX_j, Y_j) \in \mathcal{D} \times \Omega,$ $j=1,2, \ldots, n,$ be the observed data pairs, where $\bX_j=\left(\bX_{j1}^\T, \bX_{j2}^\T\right)^\T$. For a fixed $\bx_1 \in \mathbb{R}^{p_1},$ the base \F regression function is estimated by 
\begin{equation}
\label{eq:par_freg_est}
\hat{m}^{(0)}_\oplus(\bx_1)= \argmin_{\o \in \O} \sum_{j=1}^n s^{(0)}_{j}(\bx_1)d^2(\o, Y_j),
\end{equation}
where $s^{(0)}_{j}(\bx_1)=1 + (\bX_{j1}-\bar{\textbf{X}}_1)^\T \hat{\Sigma}_{11}^{-1}(\bx_1-\bar{\textbf{X}}_1)$ and $\bar{\textbf{X}}_1$ and $\hat{\Sigma}_{11}$ are the sample mean and covariance matrix of $\{\bX_{11}, \bX_{21}, \ldots, \bX_{n1}\},$ respectively. The sample partial \F $R$-squared is  {
\begin{equation}
\hat{R}^2_{\oplus,2|1} = \frac{\sum_{i=1}^{n} M_n(\hatmz{\bX_{i1}},\bX_i)- \sum_{i=1}^{n} M_n(\hatm{\bX_i},\bX_i)}{\sum_{i=1}^{n} M_n(\hatmz{\bX_{i1}},\bX_i)- \sum_{i=1}^{n} M_n(\hatm{\bX_i},\bX_i) + \sum_{i=1}^{n} d^2(\hatm{\bX_i}, Y_i)}.
    \label{eq:par_FRsquared_est}
\end{equation}}

We consider functions $g^{(0)}_{\bx_1,\o}: \mathbb{R}^{p_1} \times \O \rightarrow \mathbb{R}$ of the form $g^{(0)}_{\bx_1,\o}(\bz_1,y)= \{1 + (\bz_1 - \bm{\mu}_1)^{\mathrm{T}}\Sigma^{-1}_{11}(\bx_1 - \bm{\mu}_1) \} \{d^2(y,\o) - d^2(y, \mz{\bx_1})\},$
and the associated function classes $\mathcal{G}^{(0)}_\delta=\{g^{(0)}_{\bx_1,\o}:  \bx_1 \in \mathcal{D}_1, ~ d(\o, \mz{\bx_1}) \le \delta\},$ for $\delta > 0$. For the partial test, we impose the additional assumptions, which correspond to \ref{ass: lipschitz_metric}--\ref{ass: entropy} used for the global test.
\setcounter{assumption}{0}
\renewcommand{\theassumption}{(P\arabic{assumption})}
{\begin{assumption} \label{ass: unique_partial}
   The true and estimated base \F regressions $\mz{\bx_1}$ 
    in \eqref{eq:base_criterion} and $\hat{m}^{(0)}_\oplus(\bx_1)$ in \eqref{eq:par_freg_est} exist for all $\bx_1$, the latter almost surely, and  there exists some $\tilde{\bx}_1 \in \mathcal{D}_1$ such that  $E(d^4(\mz{\tilde{\bx}_1}, Y) ) < \infty.$
\end{assumption}
\begin{assumption}\label{ass: total_boundedness_partial}
    For all $r \in (0,1),$ there exists a totally bounded set $K_r \subset \Omega$ such that 
    \newline $\liminf_{n \to \infty} P( \hatmz{\bx_1} \in K_r \text{ for all } \bx_1  \in \mathcal{D}_1) \ge 1- r.$
\end{assumption}
\begin{assumption}\label{ass: separation_partial}
    For any $\epsilon > 0,$
    $\Delta^{(0)}(\epsilon) = \inf_{\bx_1} \inf_{d(\o,\mz{\bx_1}) > \epsilon} M^{(0)}(\o,\bx_1)-M^{(0)}(\mz{\bx_1},\bx_1) >0.$  
\end{assumption}
\begin{assumption}\label{ass: entropy_partial}
    The bracketing entropy integral  
     ${J}^{(0)}(\delta)=\int_{0}^{1} \left(1+\log N_{[]}(c \epsilon \delta , \mathcal{G}^{(0)}_\delta, L_2(P))\right)^{1/2} d\epsilon$
 satisfies $  {J}^{(0)}(\delta) = O(\delta^{-\eta}),$ for any constant $c>0,$ for some $ 0 < \eta < 1$ .
\end{assumption}}

\ref{ass: unique_partial}--\ref{ass: entropy_partial} ensure the uniqueness of the base \F regression $\mz{\bx_1}$ in \eqref{eq:base_criterion} for all $\bx_1$, and establish the uniform consistency of $\hatmz{\bx_1}$; see  Lemma~S.1 of Supplement~S.3. 

{ \begin{assumption} \label{ass: curvature}
There exist constants $0<C_1< C_2<\infty$ and sufficiently small $\delta>0$ such that $C_1 d^2(\o, \m{\bx}) \le M(\o,\bx)-M(\m{\bx},\bx) \le C_2 d^2(\o, \m{\bx}),$ for all $\bx\in\mathcal D$ and
$\omega\in B_{\delta}^{m_\oplus(\bx)}$, and $C_1d^2(\o,\mz{\bx_1}) \le M^{(0)}(\o,\bx_1) -M^{(0)}(\mz{\bx_1},\bx_1),$
for all $\bx_1\in\mathcal D_1$ and
$\omega\in B_{\delta}^{m_\oplus^{(0)}(\bx_1)}$.
\end{assumption}}


{  \ref{ass: curvature} controls the curvature of the \F  objective functions $M(\o,\bx)$ in \eqref{eq:criterion} and $M^{(0)}(\o,\bx_1)$ in \eqref{eq:base_criterion} near their respective minima. Combining with \ref{ass: lipschitz_metric}-\ref{ass: entropy} and \ref{ass: unique_partial}--\ref{ass: entropy_partial}, respectively, the lower bounds yield convergence rates for
the estimators $\hat m_\oplus(\bx)$ in
\eqref{eq:glo_freg_est} and
$\hat m_\oplus^{(0)}(\bx_1)$ in \eqref{eq:par_freg_est}.  The upper bound on the full objective, $M(\o,\bx)-M(\m{\bx},\bx)\le C_2 d^2(\o, \m{\bx}),$ is used in the  partial inference procedure but is not needed to establish convergence rates for the \F regression estimators.

\begin{remark}
    For the convergence rates for the \F regression estimators, the curvature lower bounds for the full and reduced objectives in \ref{ass: curvature} suffice. To state that result more generally, we introduce the
following weaker version of \ref{ass: curvature}.
\begin{assumptionprime}\label{ass: curvature_general}
There exist constants $C_1>0$, sufficiently small $\delta>0$, and $\alpha>1$
such that $C_1 d^{\alpha}(\o, \m{\bx}) \le M(\o,\bx)-M(\m{\bx},\bx),$ for all $\bx\in\mathcal D$ and
$\omega\in B_{\delta}^{m_\oplus(\bx)}$, and \newline $C_1d^{\alpha}(\o,\mz{\bx_1}) \le M^{(0)}(\o,\bx_1) -M^{(0)}(\mz{\bx_1},\bx_1),$ for all $\bx_1\in\mathcal D_1$ and
$\omega\in B_{\delta}^{m_\oplus^{(0)}(\bx_1)}$.
\end{assumptionprime}
\noindent Under \ref{ass: lipschitz_metric}--\ref{ass: entropy}, \ref{ass: unique_partial}--\ref{ass: entropy_partial}, and \ref{ass: curvature_general},
\begin{align*}
    & \sup_{\bx} d(\hatm{\bx},\m{\bx}) = O_P\left( \left\lbrace \frac{\log n }{n} \right\rbrace^{\frac{1}{2(\eta+\alpha-1)}}\right), \ \text{and} \ \\ & \sup_{\bx_1} d(\hatmz{\bx_1},\mz{\bx_1}) = O_P\left( \left\lbrace \frac{\log n }{n} \right\rbrace^{\frac{1}{2(\eta+\alpha-1)}}\right),
\end{align*}
where $\eta > 0$ is as  in \ref{ass: entropy} and \ref{ass: entropy_partial},
and $\alpha > 1$ as in \ref{ass: curvature_general}; for detailed proofs, see Lemma S.2 in Supplement Section S.3.
\end{remark} 
}

\begin{lemma}
	\label{lem:degeneracy_par}
	Under $H^P_0$ in \eqref{h0:partial} and \ref{ass: lipschitz_metric}--\ref{ass: entropy} and \ref{ass: unique_partial}--\ref{ass: curvature}, 
    \begin{equation*}
        \frac{1}{n}\sum_{i=1}^nM_n(\hatm{\bX_i}, \bX_i) -  \frac{1}{n}\sum_{i=1}^nM_n(\hatmz{\bX_{i1}}, \bX_{i}) = o_P(n^{-1/2}).
    \end{equation*}
\end{lemma}
As a consequence of Lemma \ref{lem:degeneracy_par}, $\sqrt{n}\hat{R}^2_{\oplus,2|1}$ is degenerate under $H^P_0$ in \eqref{h0:partial}. In analogy to the  global test, we therefore construct a test using multipliers $\phi_1, \ldots, \phi_n$ and $\psi_1, \ldots, \psi_n,$ which are independent of the data and have mean $1$ and variance $1/2,$ and the test  statistic
\begin{equation} \label{eq:test_stat_par}
T^P_n= \frac{1}{n} \sum_{i=1}^{n} \phi_i M_n(\hat{m}^{(0)}_\oplus(\bX_{i1}),\bX_i) -\frac{1}{n} \sum_{i=1}^{n} \psi_i M_n(\hatm{\bX_i},\bX_i).
\end{equation} We assume that the asymptotic variance is strictly positive  $\sigma^2_P = \frac{1}{2}E\{ M(m^{(0)}_{\oplus}(\bX_1), \bX)^2 + M(\m{\bX},\bX)^2 \}> 0.$ 
\begin{lemma}
    \label{thm:null_par}	Under $H^P_0$ in \eqref{h0:partial} and \ref{ass: lipschitz_metric}-\ref{ass: entropy} and \ref{ass: unique_partial}--\ref{ass: curvature}, as $n \rightarrow \infty$,
	\begin{equation*}
	\frac{n^{1/2}T^P_n}{\hat{\sigma}_P} \rightarrow N(0,1),
	\end{equation*}
	where $\hat{\sigma}_P^2= \frac{1}{2n} \sum_{i=1}^{n}  \{ M_n(\hat{m}^{(0)}_\oplus(\bX_{i1}),\bX_i)^2 + M_n(\hatm{\bX_i},\bX_i)^2  \}$ is a consistent estimator of $\sigma^2_P.$
\end{lemma}

Again adopting \eqref{eq:test_stat_Cauchy}, we obtain the Cauchy combination test statistic
\begin{equation}
     C^{P}_n = \sum_{k=1}^{K} c_{k} \text{tan}\{(2\Phi(\lvert \tilde{T}^{P}_{n,k} \rvert) - 3/2)\pi\},
    \label{eq:test_stat_Cauchy_partial}
\end{equation} where $(\tilde{T}^{P}_{n,1}, \ldots, \tilde{T}^{P}_{n,K})$ is a collection of $K$ randomized test statistics $n^{1/2}T^P_n/ \hat{\sigma}_P.$

\begin{theorem}
    \label{thm:null_par_cauchy}
   Under $H^P_0$ in \eqref{h0:partial} and \ref{ass: lipschitz_metric}--\ref{ass: entropy} and \ref{ass: unique_partial}--\ref{ass: curvature}, for the test statistic $C^{P}_n$ in \eqref{eq:test_stat_Cauchy_partial}, as $n \rightarrow \infty$, $C^{P}_n \rightarrow \text{Cauchy}(0,1)$ in distribution.
   \end{theorem}

As before,  the rejection region for $H^P_0$ in \eqref{h0:partial} at level $\alpha$ is {$\mathcal{R}_{n,P}= \left \lbrace  C^{P}_n   > \rho_{\alpha} \right \rbrace,$} where $\rho_{\zeta}$ is the upper $\zeta$-quantile of \textit{Cauchy}$(0,1)$.  We consider {local} alternatives \begin{align}
    { H^{P}_{1,n} = \{Q : D_{2\mid1}(Q) \ge b_n \},}
    \label{eq:contig_alt_par}
\end{align} with $D_{2\mid1}(\cdot)$ in \eqref{eq:numerator_diff_partial}. 
We assume that $\Sigma=\operatorname{var}(\bX)$  is uniformly positive definite for all $n$ and $Q \in H^{P}_{1,n}.$  The corresponding partial test has power $\beta_n^P
    = Q \left( \mathcal{R}_{n,P} \right).$

\begin{theorem} \label{thm: power_partial}
Assume \ref{ass: lipschitz_metric}--\ref{ass: entropy} and \ref{ass: unique_partial}--\ref{ass: curvature} hold uniformly in $Q \in H^{P}_{1,n}$  in \eqref{eq:contig_alt_par}, and let $b_n$ be a sequence such that $b_n \to 0$ and $n^{1/2}b_n \to \infty.$ Then, one has power $\inf_{Q\in H^P_{1,n}}\beta^{P}_{n}(Q) \to 1.$ 
\end{theorem}

\section{Implementation and Simulations}\label{sec:simul}

\subsection{General approach}

To evaluate the finite sample behavior of the global and partial \F regression test, simulations with network-valued and compositional-valued responses are presented in Section \ref{sec:simul:network} and \ref{sec:simul:compositional}, respectively. 
The number of Cauchy combination components in test statistics is set to $K = 100$, with equal weights $c_k = 1/K$ in \eqref{eq:test_stat_Cauchy} and \eqref{eq:test_stat_Cauchy_partial} (see Remark 2 in \citet{liu2020cauchy}). The multipliers $\psi_i$ and $\phi_i$ used in constructing the test statistics are generated from $N(1, 1/2).$ To illustrate empirical power of the tests, we implement $B = 1000$ Monte Carlo replications with significance level $\alpha = 0.05.$ Further analysis on the effect of $K$ can be found in Supplement Section S.2.3.




\subsection{Network simulation}\label{sec:simul:network}

We focus on $m \times m$ graph Laplacian responses $G$ in the space of networks $\mathbb{L}^{m\times m}$ with the Frobenius metric $d$ in Example 1 of Section \ref{sec:prelim}, with three-dimensional predictors $\bX = (X_1, X_2, X_3).$ For notational convenience, let $\operatorname{vech}(A)$ be the vectorization of the strictly lower triangular part of symmetric $m \times m$ matrix $A,$ yielding a vector of length $q = m(m - 1)/2$. Given the symmetry and zero row-sum properties of Laplacians, any graph Laplacian $G$ is fully characterized by vech$(G)$. The inverse operation vech$^{-1}$ maps such vectors back to Laplacian matrices.


We first generate predictors $\textbf{X} = (X_{1},X_{2}, X_3)$ and a latent random effect $W$ such that  $(X_{1},X_{2}, X_3, W) = (\Phi(V_{1}),\Phi(V_{2}),\Phi(V_{3}),\Phi(V_{4})) \in [0,1]^4,$ where $(V_{1}, V_{2}, V_{3}, V_4)^{\T} \sim N_{4}(0, \Sigma)$ with $\Sigma = (a_{kl})$ such that $a_{kk} = 1$ and $a_{kl} = 1/2,$ for $k \neq l,$ and $\Phi$ is the standard normal cumulative distribution function. We denote $m_{\oplus, \mathrm{GL}}(\bx)$ as the \F regression function of graph Laplacian responses $G$ given $\bx = (x_1, x_2, x_3) \in [0,1]^3.$ The purpose of latent random effect $W$ is to induce dependency between the predictors and the response even when there is no regression effect, as detailed in the following.

For network responses, we consider a weighted stochastic block model \citep{aicher2015learning} with two communities defined by membership vectors $\textbf{z} = (\textbf{z}^{\T}_1, \textbf{z}^{\T}_2)^{\T}$, where each $\textbf{z}_i$ is an $m_i$-dimensional vector with identical elements $i$, and $m_1 + m_2 = m$. Then the graph Laplacian response from the stochastic block model is represented as a block matrix $G = (G_{kl}),$ where each submatrix $G_{kl} \in \real^{m_{k} \times m_{l}},$ for $k,l=1,2,$ corresponds to the connections within or between the $k$th and $l$th communities. The existence of an edge between nodes of each block $G_{kl}$ is governed by  a Bernoulli distribution with success probability $p_{kl}.$ We set $p_{11} = p_{22} = 0.9$ and $p_{12} = p_{21} = 0.7$ and  
 generate random network responses as
\begin{equation*}
    G = \operatorname{vech}^{-1}[-\operatorname{vech}(B)  \odot (g_1, \ldots, g_q) ],
\end{equation*}
where $\odot$ denotes the Hadamard (element-wise) product, symmetric block matrix of edge indicators $B=(B_{kl})_{k,l=1}^{2}$ with each element in the $m_k \times m_l$ matrix $B_{kl}$ following a Bernoulli($p_{kl}$) distribution for $k, l = 1, 2,$ and each $g_j,$ $j = 1, \ldots, q,$ sampled from 
\begin{equation}
    g_{j} | \bX, W \sim \text{Unif}\left(\beta_0 + \beta_1 X_1 + \beta_2 X_{2} + \beta_3 X_{3} - W,~  \beta_0 + \beta_1 X_1 + \beta_2 X_{2} + \beta_3 X_{3} + W \right),
    \label{eq:datgen_den}
\end{equation} where $\beta_1,$ $\beta_2$ and $\beta_3$ are effect size parameters. 

 Under this construction, the pair $(\bX, G)$ satisfies the global \F regression model
\begin{equation}
    m_{\oplus, \mathrm{GL}}(\textbf{x}) = \operatorname{vech}^{-1}[ -\operatorname{vech}(\Lambda) \odot (s(\textbf{x}), \ldots, s(\textbf{x}))],\quad s(\bx) = \beta_0 + \beta_1 x_1 + \beta_2 x_2 + \beta_3 x_3,
    \label{eq:simul_net_pop}
\end{equation} where $\Lambda$ is an $m \times m$ matrix composed of probability parameter block matrices $(p_{kl})_{m_k \times m_l}$ with all entries equal to $p_{kl}$, for $k,l = 1,2$; see Section 5.3 of \cite{zhou2022network}. 
 
 Note that the $W$ in \eqref{eq:datgen_den} affects the variability of each response element $g_j$ but not the regression function $m_{\oplus, \mathrm{GL}}(\bx)$ in \eqref{eq:simul_net_pop}.
This setting emphasizes the distinction between testing for independence and testing for a \F regression effect: independence implies the absence of a regression effect, however the converse is not true, and in the above setting independence is violated,  yet there is no regression effect. 

When $\beta_1 = \beta_2 = \beta_3 = 0,$ $m_{\oplus, \mathrm{GL}}(\bx) = \operatorname{vech}^{-1}[ -\operatorname{vech}(\Lambda) \odot (\beta_0, \ldots, \beta_0)]$ in \eqref{eq:simul_net_pop} is constant in $\bx,$ corresponding to the null hypothesis $H^{G}_{0}$ in \eqref{h0:global} of no regression effect. 
To evaluate the empirical performance of the proposed global \F regression test, we consider the null hypothesis $H^{G}_0: m_{\oplus, \mathrm{GL}}(\bx) = \omega_{\oplus}$ in \eqref{h0:global} by setting $\beta_0 = 1,$ and $\beta_1 = \beta_2 = \beta_3 = 0,$ and alternatives $H^{G}_1$, with increasing $0 < \beta_1 = \beta_2 = \beta_3 \le 0.2.$

To illustrate the performance of the partial \F regression test, we consider a setting in which the base model contains one predictor $X_1,$ and we examine whether the additional predictors $X_2$ and $X_3$ contribute to the model. When $\beta_1 > 0$ and $\beta_2 = \beta_3 = 0,$ the regression function
$m_{\oplus, \mathrm{GL}}(\bx) = 
\operatorname{vech}^{-1}[ -\operatorname{vech}(\Lambda) \odot (\beta_0 + \beta_1 x_1, \ldots, \beta_0 + \beta_1 x_1)]$ in \eqref{eq:simul_net_pop} depends only on $x_1,$ satisfying the null $H^{P}_0: m_{\oplus, \mathrm{GL}}(\bx) =  m^{(0)}_{\oplus, \mathrm{GL}}(x_1)$ in \eqref{h0:partial}. Under this null simulation $H^{P}_0$, we set $\beta_0 = 1,$ $\beta_1 = 0.25,$ and $\beta_2 = \beta_3 =0$.
To evaluate empirical power, we consider alternatives $H^{P}_{1}$ by fixing $\beta_1 = 0.25$ and increasing $0 < \beta_2 = \beta_3 \le 0.35.$ 



\begin{figure}[h!]
    \centering
    \includegraphics[width=0.35\linewidth]{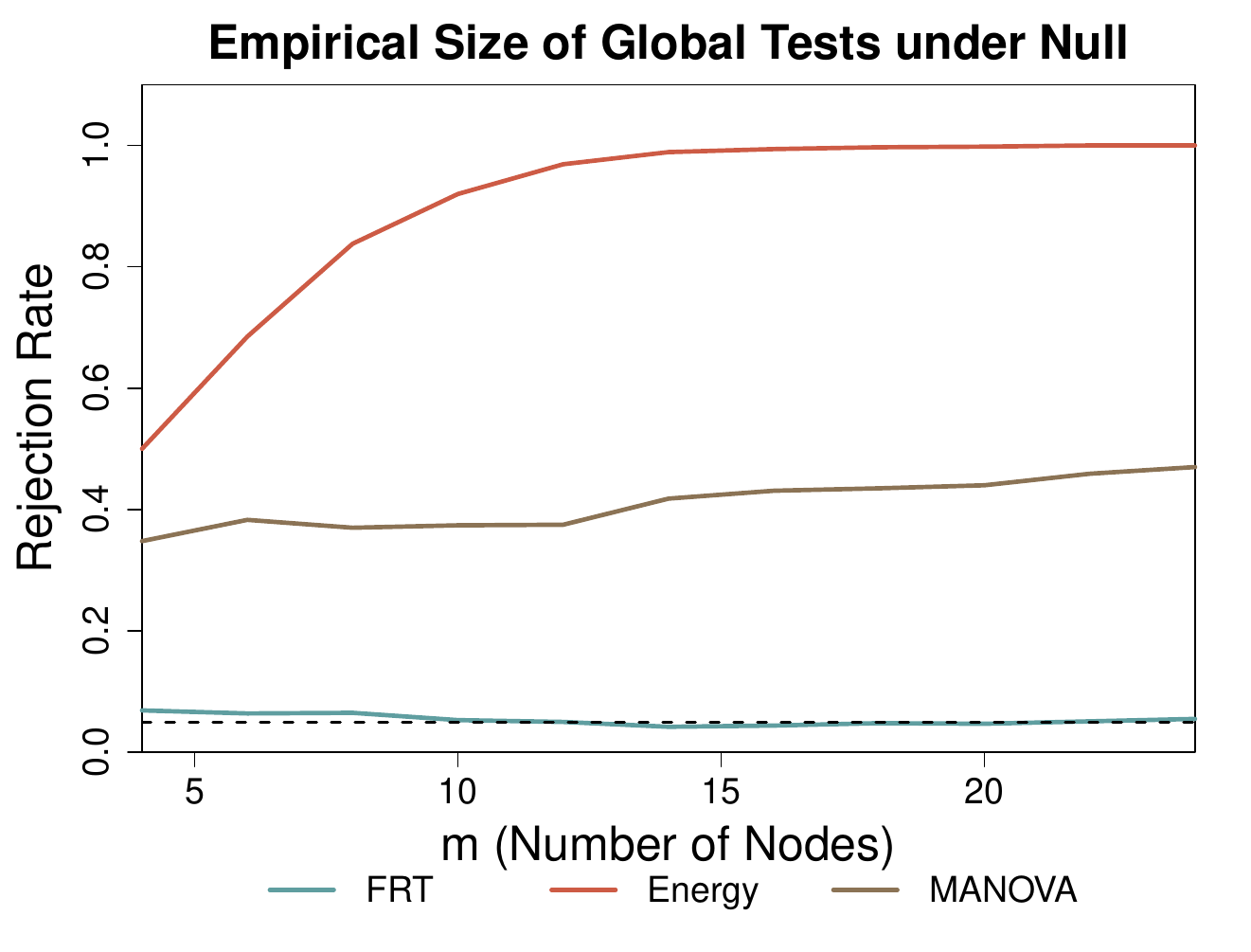}
    \includegraphics[width=0.35\linewidth]{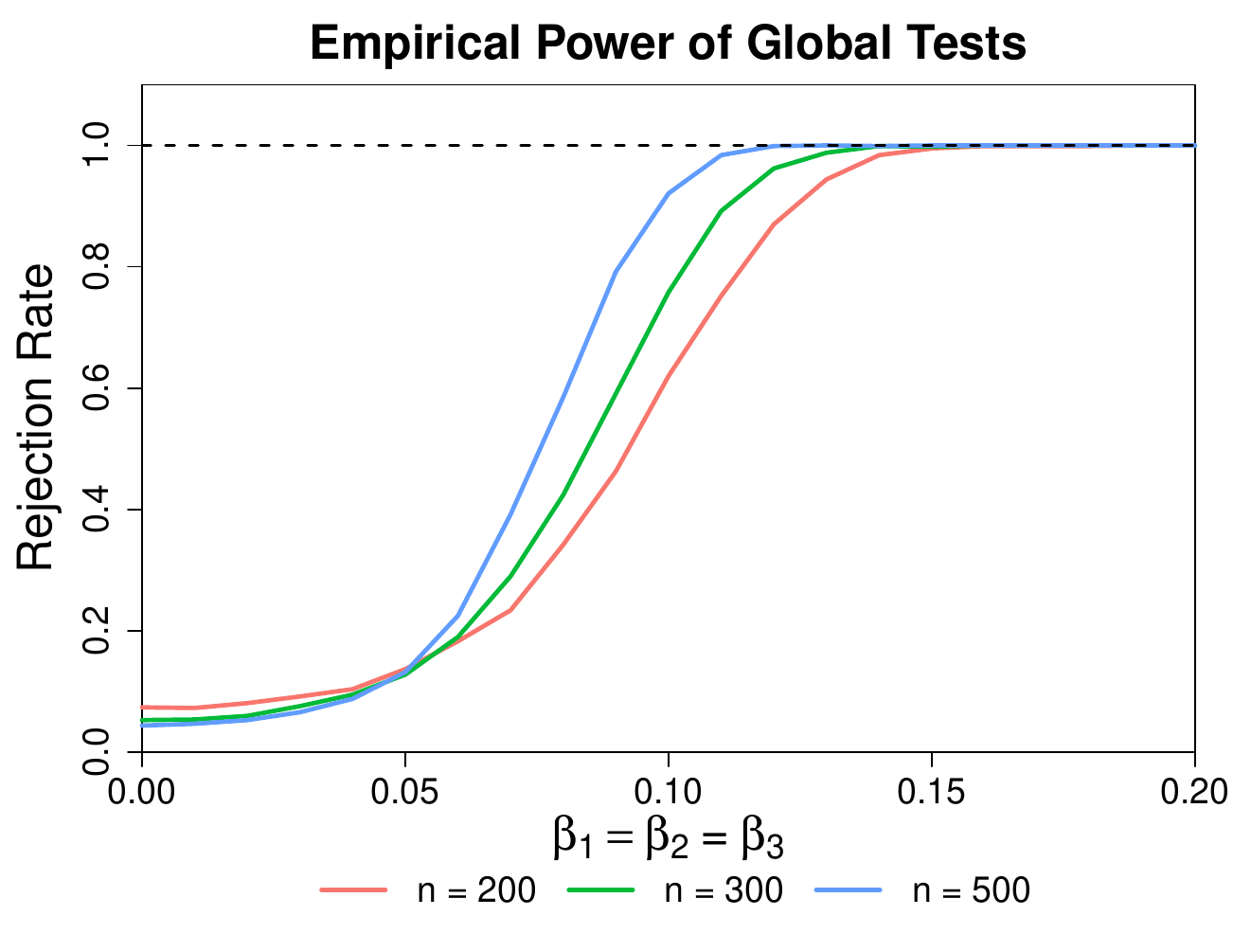}
    \includegraphics[width=0.35\linewidth]{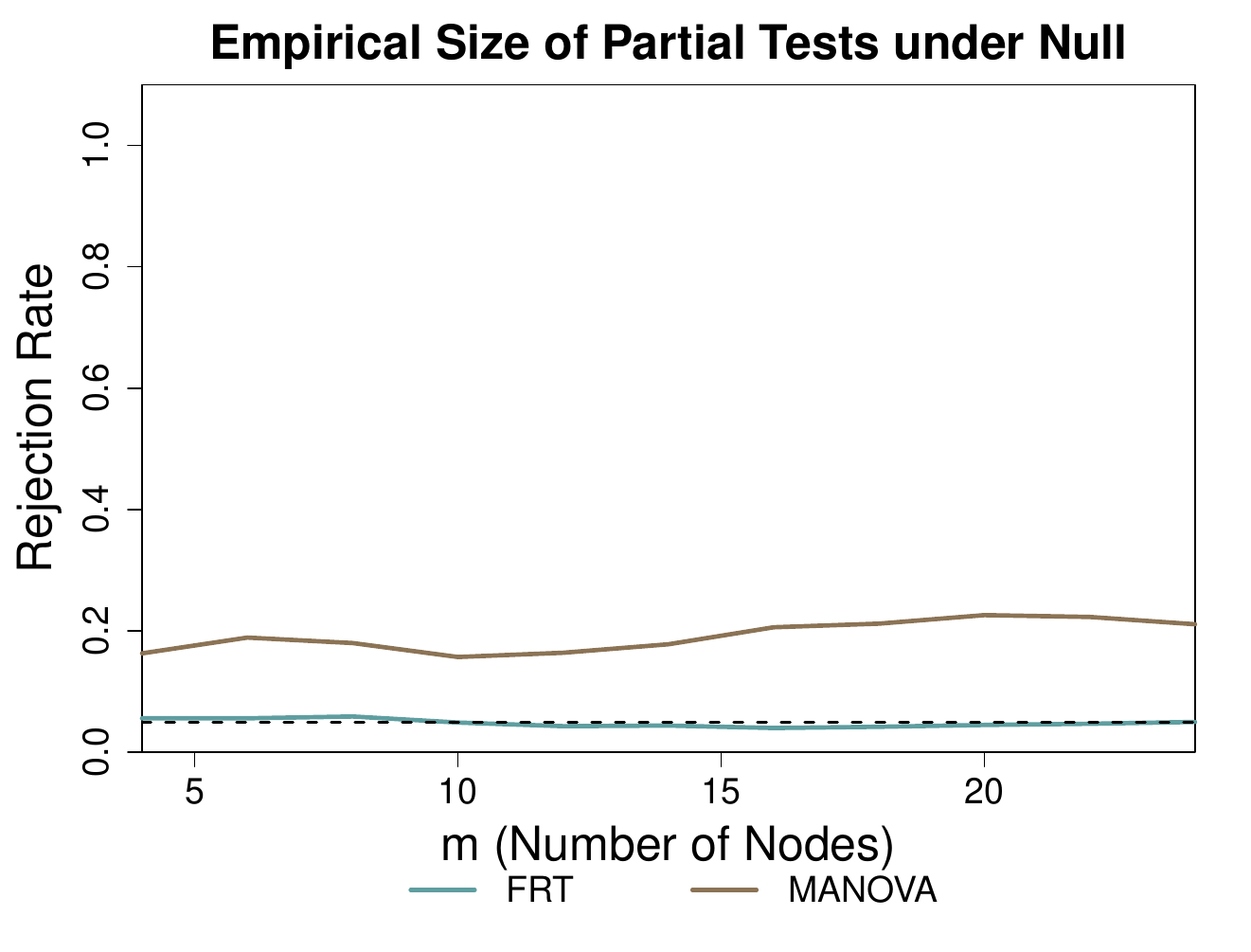}
    \includegraphics[width=0.35\linewidth]{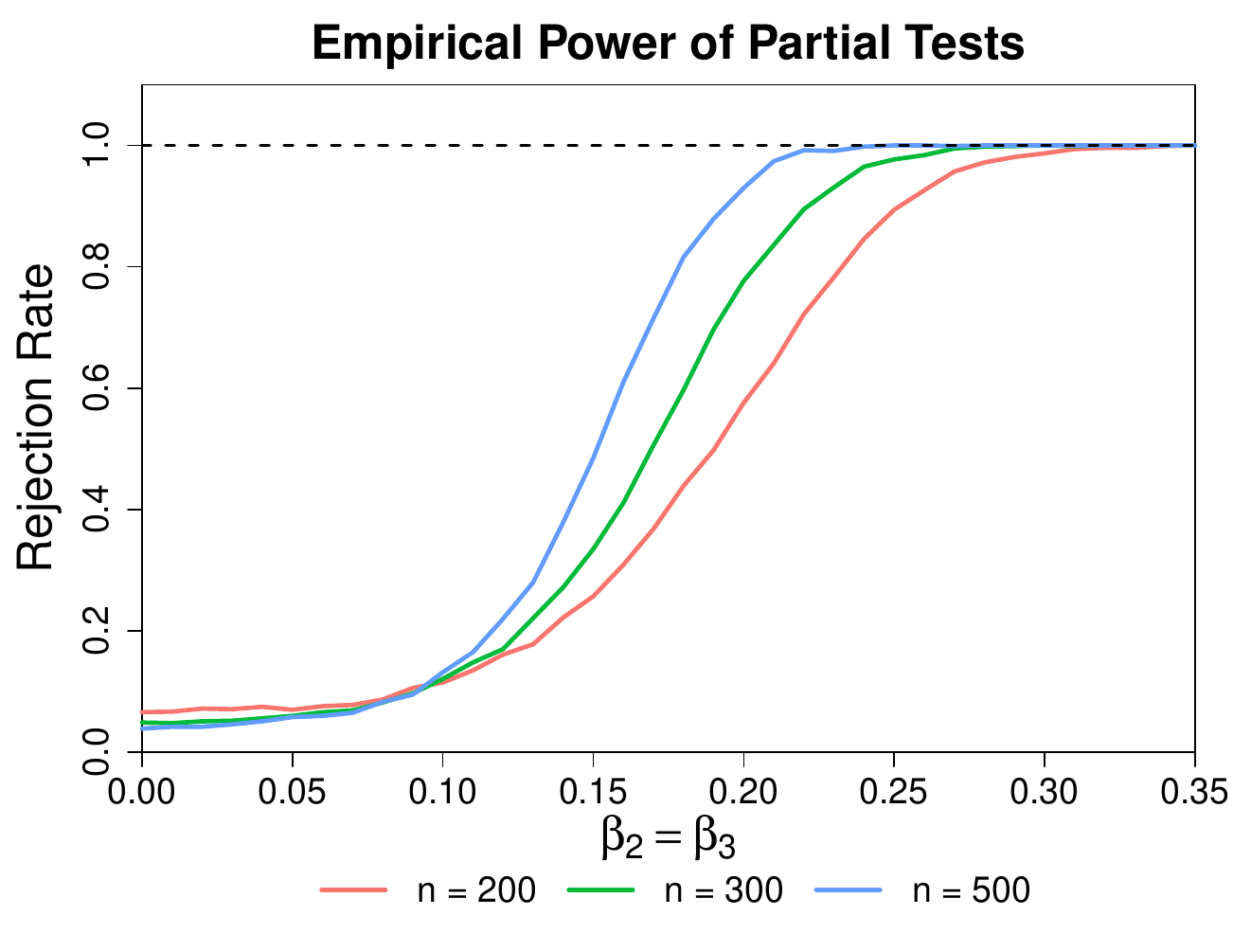}
    \caption{Finite sample performance of network response simulations with $\alpha =0.05$. {The methods compared include the \F regression test (FRT),  energy distance independence test (Energy), and multivariate analysis of variance (MANOVA) test on $\operatorname{vech}(G) \in \mathbb{R}^{m(m-1)/2}.$}
    (Top left) Empirical rejection rates under the global null $H^{G}_0$  against the number of nodes $m$ with  $n = 300.$ (Top right) Empirical power of the FRT under the global alternative $H^{G}_1$ against increasing  effect sizes, where  $0 \le \beta_1=\beta_2=\beta_3 \le 0.2$ for $n = 200, 300$ and $500.$ (Bottom left) Empirical rejection rates under the partial null $H^{P}_0$ with $n = 300$. (Bottom right) Empirical power of FRT under the partial alternative $H^{P}_1$ against the effect size $0 \le \beta_2 = \beta_3 \le 0.35,$ for $n = 200, 300$ and $500.$}
    \label{fig:simul_net_glo}
\end{figure}

For global tests, we compare three methods:  The \F regression test (FRT),   the energy distance independence test (Energy) \citep{szekely2007measuring}, and {Multivariate Analysis of Variance (MANOVA) on the vectorized strictly lower triangle of graph Laplacian $\operatorname{vech}(G) \in \mathbb{R}^{m(m-1)/2}.$}
The first row of Figure \ref{fig:simul_net_glo} demonstrates empirical rejection rates under the global null $H^{G}_0$ and alternative $H^{G}_{1}$ with significance level $\alpha = 0.05.$ Under $H_0^{G},$ we generate $n = 300$ independent samples of $(\bX_i, G_i),$ $i = 1, \ldots, n,$ and vary the number of nodes from $m = 2$ to $m = 24.$ Here, the energy test rejects the null hypothesis even when $\beta_1 = \beta_2 = \beta_3 = 0$,  due to the dependence between $\bX$ and $Y$ that arises  from the inclusion of the latent effect $W$ 
in \eqref{eq:datgen_den}. 
However, the  presence of $W$ does not affect the underlying conditional \F regression  $m_{\oplus, \mathrm{GL}}(\bx)$, and this is reflected by the performance of the FRT, which maintains the nominal significance level $\alpha = 0.05$ across different numbers of nodes $m$ in the network response $G.$ The MANOVA test does not preserve the nominal significance level $\alpha$ under the null. Under $H^{G}_{1},$ we examine the empirical power of the FRT with sample sizes $n = 200, 300,$ and $500.$ The right panel in the first row shows that the empirical power of the FRT increases with the effect size of $\beta_1,$ $\beta_2,$ and $\beta_3$ with larger sample sizes yielding higher empirical power, as expected. 


For the partial testing, we compare two methods, FRT and MANOVA.
The second row of Figure \ref{fig:simul_net_glo} displays  rejection rates under the partial null $H^{P}_0$ and alternative $H^{P}_{1}.$ Under $H^{P}_0,$ the left  panel in the second row shows that the FRT maintains $\alpha = 0.05$ across different numbers of nodes $m$ with  $n = 300,$ whereas the MANOVA test does not maintain the nominal level $\alpha.$ Under $H^{P}_{1},$ the right panel in the second row shows that the empirical power of the FRT increases to $1$ as the partial effect size $\beta_2 = \beta_3 > 0$ increases, with larger sample sizes leading to better power  performance.

\subsection{Compositional data simulation}\label{sec:simul:compositional}

We consider compositional responses $Y = (Y_1, Y_2, Y_3)$ in the $2$-dimensional simplex $\Delta^{2}$, transformed via taking component-wise square roots to the positive orthant of the unit sphere $U  = (U_1,U_2,U_3) = (\sqrt{Y_1},\sqrt{Y_2},\sqrt{Y_3}) \in \mathbb{S}_{+}^{2}$ with the geodesic spherical distance $d$ (see Example 2 of Section~\ref{sec:prelim}) 
and pair these with two dimensional predictors $\bX = (X_1, X_2) \in \left[0,1\right]^2,$ with $X_j\stackrel{iid}{\sim} \text{Unif}[0,1],$ for $j = 1, 2.$ Define the latent variable $W \sim \text{Unif}[0, a(\bX)],$ where $a(\bx) = 1/2 + 1/4\{(2x_1 -1)^2 + (2x_2 -1)^2 \}.$ 
Under this setting,  each of  $X_1$ and $X_2$ is dependent on $W.$ Following the previous network simulation, the latent effect $W$ induces dependence between the predictors and the response even in the absence of a regression effect, as detailed below.

We define $u(\bx) = \left(\sin(h(\bx)), \cos (h(\bx))\cos(\phi), \cos (h(\bx))\sin(\phi)\right),$ with $h(\bx) = \alpha_0 + \beta x_1 + \gamma x_2,$ where $0 <\phi <\frac{\pi}{2}$ is a fixed rotational angle, and $\beta, \gamma \ge 0$ are effect size parameters with constraints $\alpha_0 > 0$ and $0 < \alpha_0 +\beta + \gamma < \pi/2$ to ensure spherical responses lie in the positive orthant of the sphere. We generate spherical responses $U$ by introducing a random perturbation
in the tangent space at $u(\bX),$ i.e., $T_{u(\bx)} \mathbb{S}^2_+ = \{ v \in \mathbb{R}^3 \mid \langle v, u(\bx) \rangle = 0 \},$ and mapping it onto the unit sphere
via the Riemannian exponential map $\operatorname{Exp}_{u(\bx)}.$ Conditional on $\bX$, the spherical data $U$ is generated as \[
U \mid \bX = \text{Exp}_{u(\bX)}(\mathbf{e}) = \cos(\lVert \mathbf{e}  \rVert){u(\bX)} + \sin(\lVert \mathbf{e} \rVert)\frac{\mathbf{e} }{\lVert\mathbf{e}  \rVert},
\]
where $\mathbf{e} = W( e_1\nu_1 + e_2\nu_2),$ with $e_j = \sigma(2Z_j -1),$ $Z_j \stackrel{iid}{\sim}  \text{Beta}(1/2, 1/2),$ $j = 1, 2,$ and  $\nu_1 = (0, -\sin(\phi), \cos(\phi))$ and $\nu_2 = (\cos(h(\bx)), -\sin(h(\bx))\cos(\phi), -\sin(h(\bx))\sin(\phi))$ form an orthonormal basis of the tangent space $T_{u(\bx)}\mathbb{S}^{2}_{+}.$ We choose $\sigma > 0$ sufficiently small so that $\lVert \mathbf{e} \rVert$ lies in the injectivity radius of $\mathbb{S}^2_{+},$ ensuring that $\text{Exp}_{u(\bX)}(\mathbf{e})$ is well defined and yields a small geodesic perturbation around $u(\bX).$


For the global test,
we set $\alpha_0 = \pi/6,$ $\beta = \gamma = 0$ under $H^{G}_0,$ and consider $0 < \beta = \gamma \le  0.08$ under alternatives $H^{G}_1.$  
For the partial \F test, we consider a base predictor $X_1,$ and examine whether the additional predictor $X_2$ contributes to the model. We set 
$\alpha_0 = \pi/6,$ $\beta = \pi/12,$ and $\gamma = 0$ 
under this null $H^{P}_0$, and consider $0 < \gamma \le  0.14$ under the alternative $H^{P}_{1}.$ 
The rotation angle and noise level remain $\phi = \pi/4$ and $\sigma = 0.15.$

\begin{figure}[!h]
    \centering
    \includegraphics[width=0.35\linewidth]{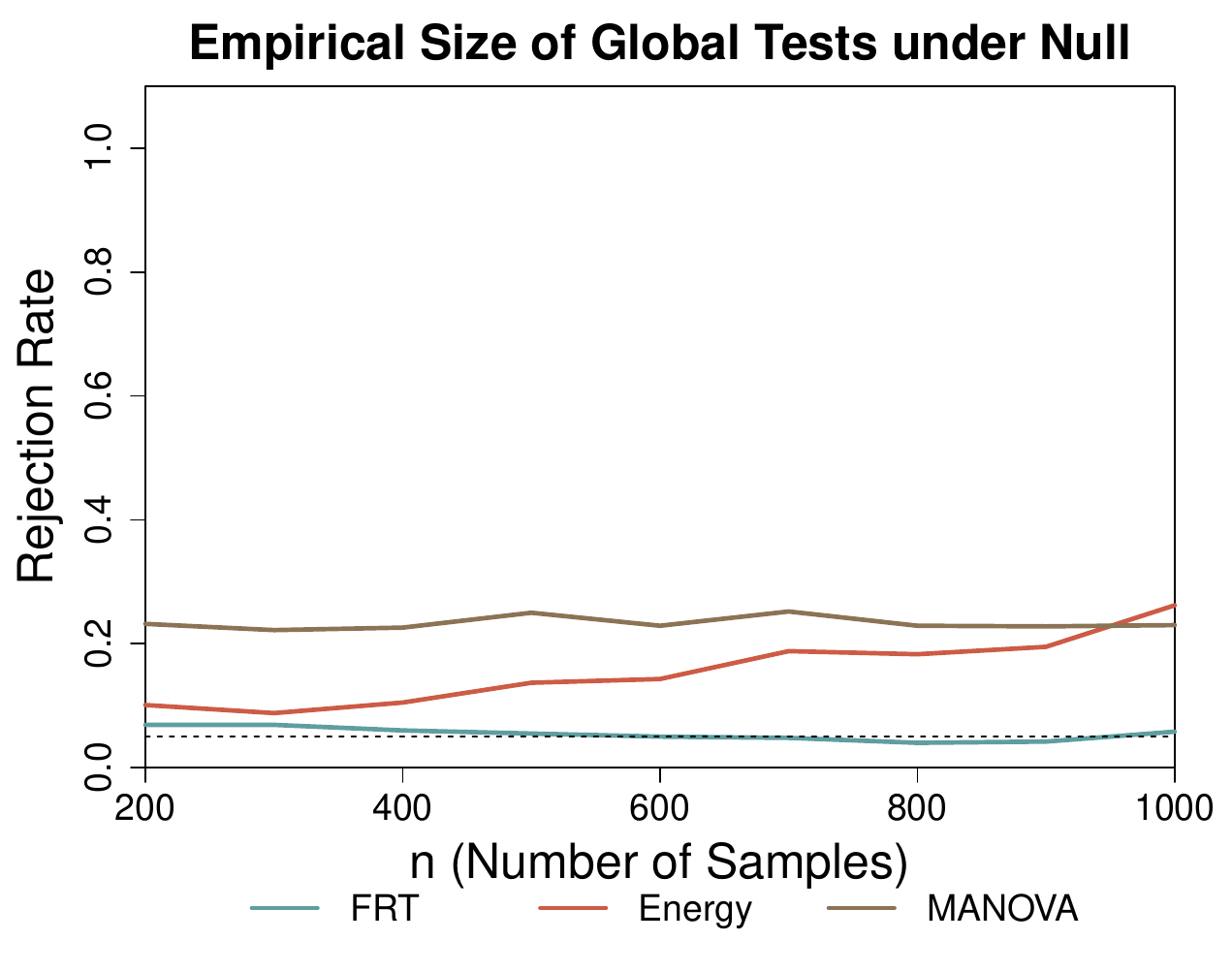}
    \includegraphics[width=0.35\linewidth]{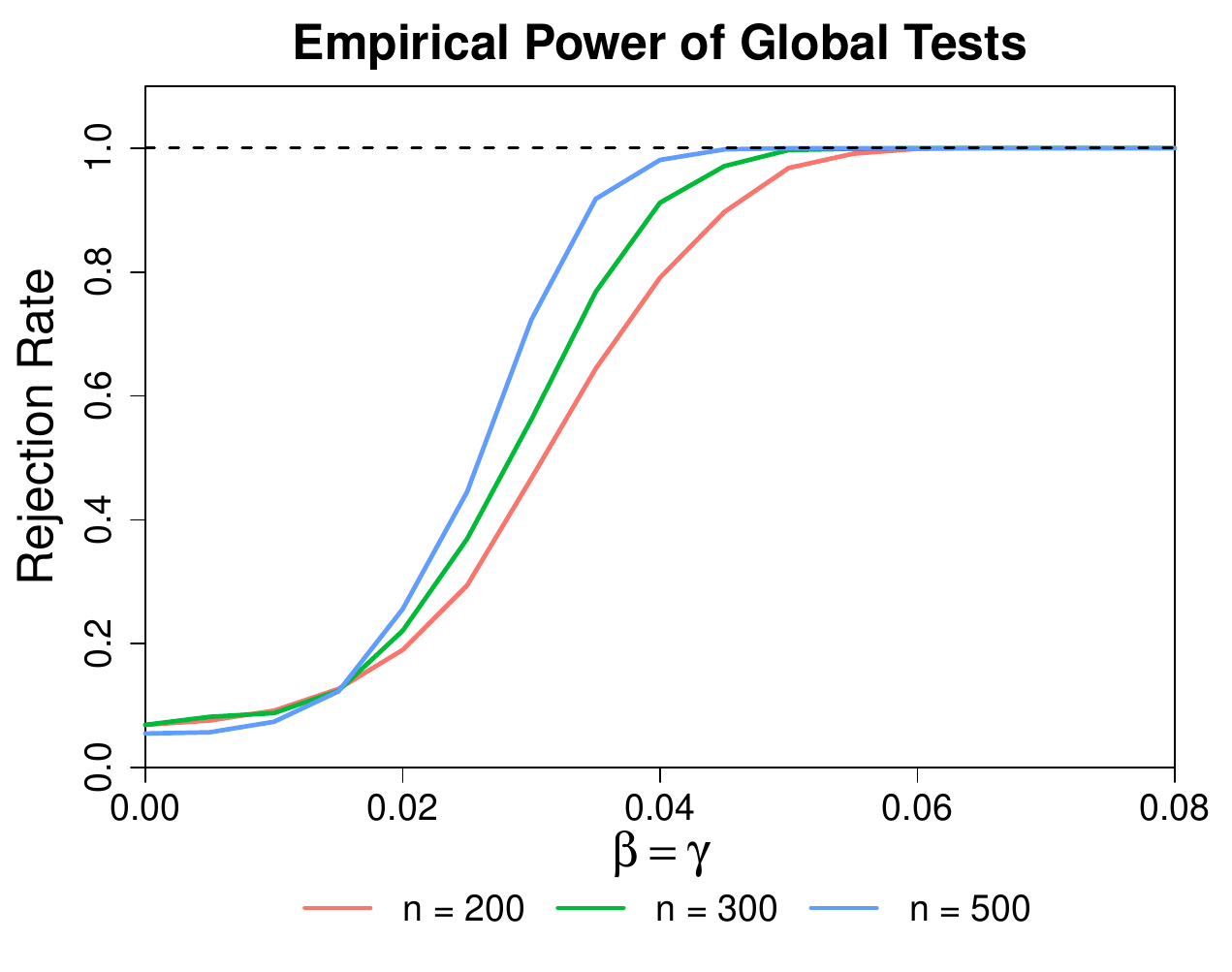}
    \includegraphics[width=0.35\linewidth]{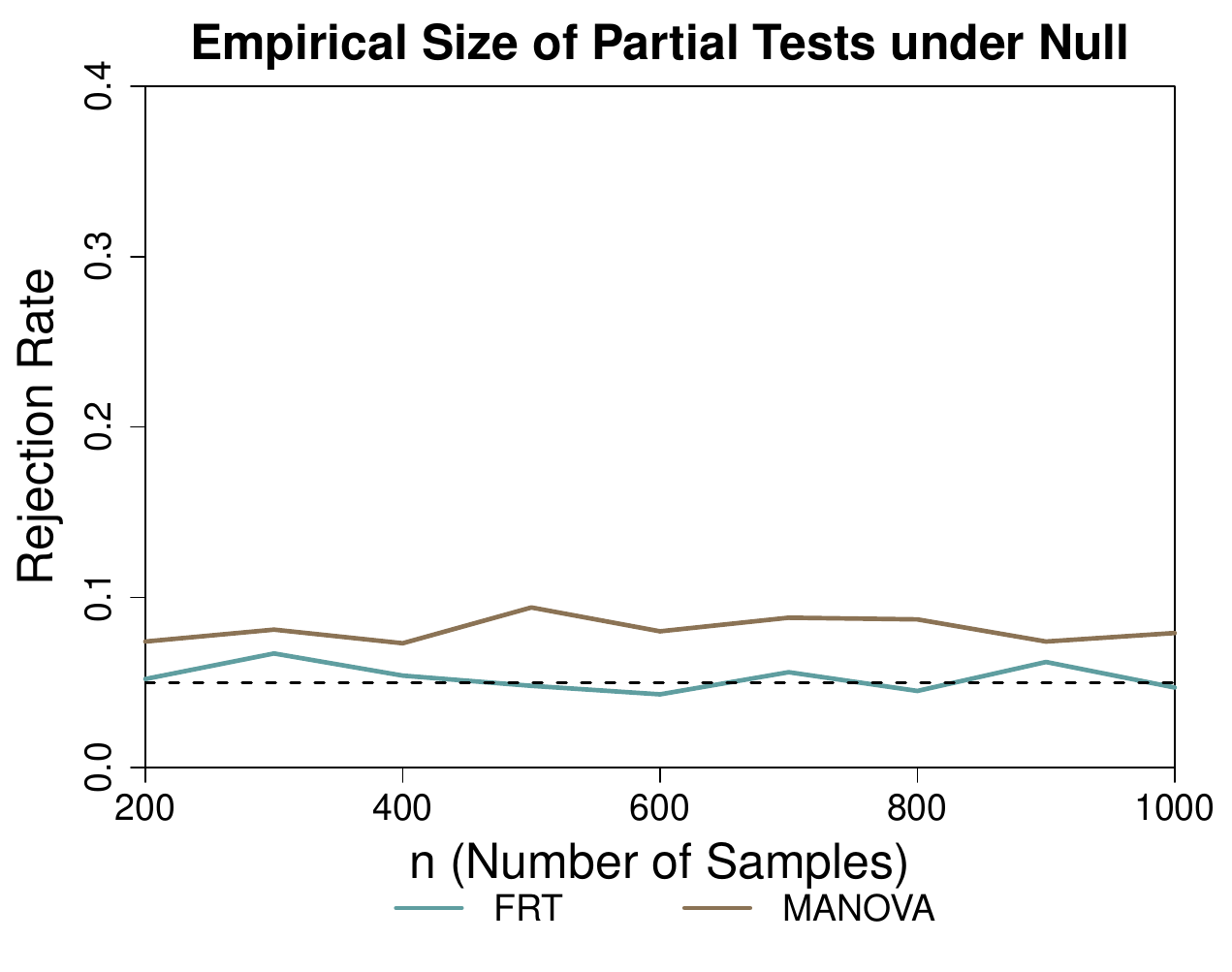}
    \includegraphics[width=0.35\linewidth]{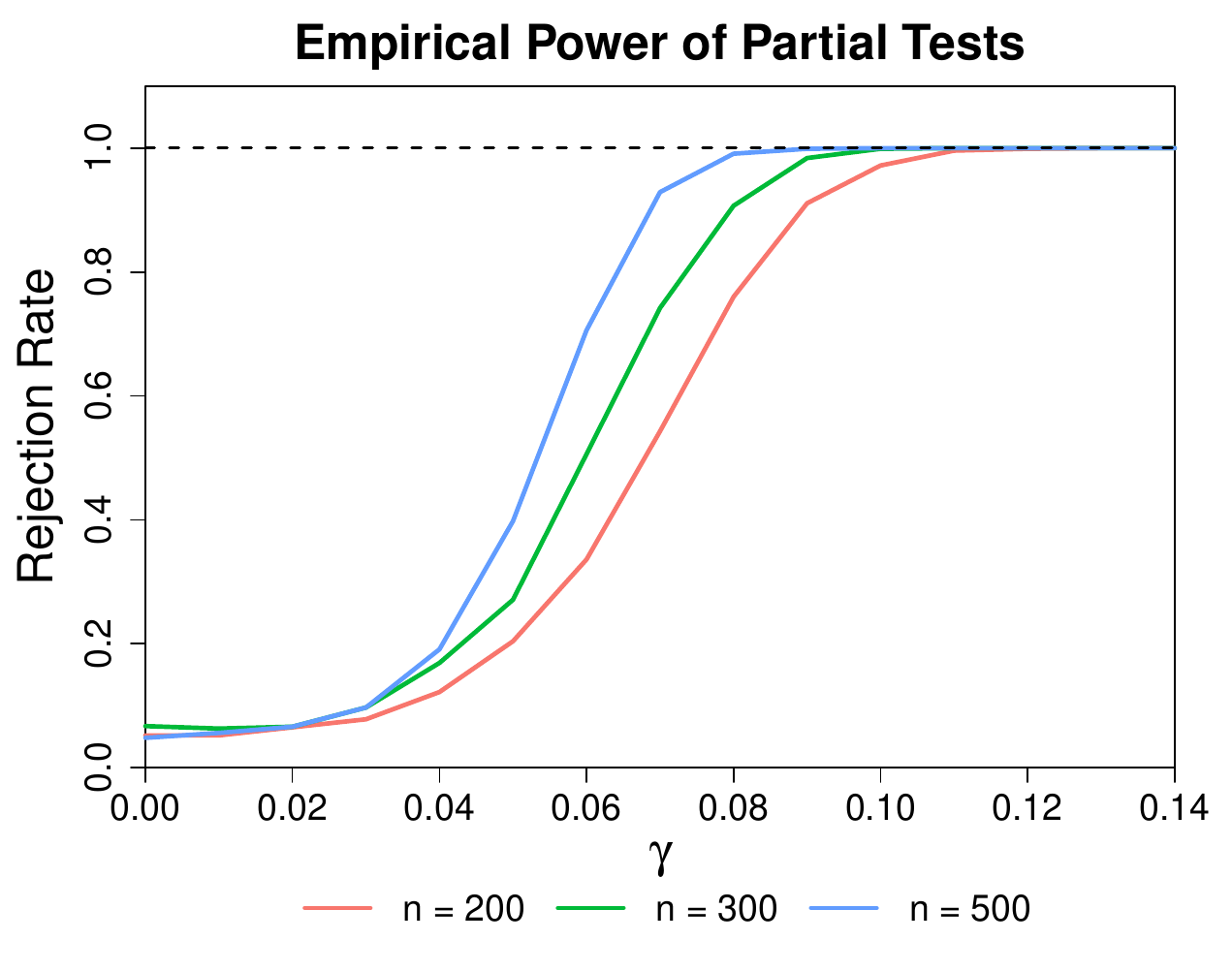}
    \caption{Finite sample performance of compositional response simulations with $\alpha =0.05$. The methods compared include the \F regression test (FRT), energy distance independence test (Energy), and multivariate analysis of variance (MANOVA) test on isometric log-ratio coordinates of the centered log-ratio transformed compositional responses $Y.$ (Top left) Empirical rejection rates under the global null $H^{G}_0$  against sample sizes $n.$ (Top right) Empirical power of the FRT under the global alternative $H^{G}_1$ against the effect sizes $\beta$ and $\gamma,$ for  $n = 200, 300$ and $500.$ (Bottom left) Empirical rejection rates under the partial null $H^{P}_0$ over $n$ and (Bottom right) Empirical power of FRT under the partial alternative $H^{P}_1$ against the effect sizes $\gamma,$ for $n = 200, 300$ and $500.$}
    \label{fig:simul_sph}
\end{figure}

We compare three methods for global testing: the proposed FRT, the energy test \citep{szekely2007measuring}, and a {MANOVA test on isometric log-ratio coordinates obtained by projecting the centered log-ratio transforms of the compositional responses onto an orthonormal basis} \citep{aitchison1982statistical,egozcue2003isometric}.
The first row of Figure \ref{fig:simul_sph} shows empirical rejection rates under the global null $H^{G}_0$ and alternative $H^{G}_{1}$ at significance level $\alpha = 0.05$. Under $H_0^{G},$ where $\beta = \gamma = 0,$ the energy test shows inflated rejection rates due to the dependence induced by the $W$ term through the perturbation $\mathbf{e}.$ In contrast, 
the FRT maintains the level $\alpha = 0.05$ across different sample sizes up to $n = 1000.$ The MANOVA test does not preserve its nominal level $\alpha$ under the null. 
Under $H^{G}_{1}$ in the right panel, the empirical power of FRT increases with the effect sizes of $\beta$ and $\gamma$ and approaches $1,$ with larger sample sizes yielding higher empirical power.

For partial testing, we compare the FRT and MANOVA test. The second row of Figure \ref{fig:simul_sph} presents rejection rates under the partial null $H^{P}_0$ and alternative $H^{P}_{1}.$ Under $H^{P}_0,$ FRT maintains the level $\alpha = 0.05$ across sample sizes $n.$ The MANOVA method shows increasing rejection rates as $n$ grows. 
Under $H^{P}_{1},$ the empirical power of the FRT approaches $1$ as the partial effect size $\gamma > 0$ increases.

\section{Data Illustrations}\label{sec:data}

\subsection{New York City taxi networks}\label{sec:taxi}

We apply the proposed \F regression inference to New York City (NYC) yellow taxi trip data, which are publicly available at \url{https://www.nyc.gov/site/tlc/about/tlc-trip-record-data.page}. These data contain trip-level information, including pick-up and drop-off dates and locations, trip distances, payment types, and passenger counts. Random objects of interest are daily transport networks of passenger movement across regions in Manhattan, restricting the analysis to the period from January 1, 2017 to December 31, 2019, to avoid COVID-19 pandemic effects. Following the preprocessing  in \cite{zhou2022network}, the 66 original taxi zones are aggregated into 13 regions within Manhattan; see Appendix D of \cite{zhou2022network} for additional details.
For each day, we construct an undirected transport network in which nodes correspond to the 13 regions and edge weights represent the daily volume of passenger flow between pairs of regions. Each network is represented by its associated $13 \times 13$ graph Laplacian matrix, as described in Example 1 of Section~\ref{sec:prelim}, and these graph Laplacians are the response objects. 
Predictors include daily trip-level features aggregated at the day level, including the average trip distance ($X_1$) and average fare per trip ($X_2$). We also incorporate day of week information, encoded through indicator variables for Wednesday-Thursday ($X_3$), Friday ($X_4$), Saturday ($X_5$), and Sunday ($X_6$), respectively, with Monday-Tuesday as the baseline.

We first conduct a global \F regression test to assess whether the predictor vector $\bX = (X_1, \ldots, X_6)$ is associated with the transport network response $Y$. To further examine variable-specific contributions, we perform partial \F regression tests. For each predictor $X_j$, we partition the predictors into $\bX_1 = \bX \setminus X_j$ and $\bX_2 = X_j$, and test whether inclusion of $X_j$ provides additional explanatory power beyond $\bX_1$. For test statistics $C_G$ in \eqref{eq:test_stat_Cauchy} and $C_P$ in \eqref{eq:test_stat_Cauchy_partial}, we use the number of Cauchy combination components $K = 100$ with equal weights $c_k = 1/K,$ following the simulation setting in Section \ref{sec:simul}. 

Table \ref{tab:Taxi_pvalues} presents the resulting $p$-values.  
For the global test, the small $p$-value provides strong evidence against the global null hypothesis of no regression effect. For the partial tests, the fare variable $X_2$ and the Wednesday--Thursday indicator $X_3$ do not provide significant additional explanatory power when conditioning on the remaining predictors at significance level $\alpha = 0.05.$ This may be attributed to correlations among predictors, especially between trip distance $X_1$ and fare $X_2$, and between Wednesday-Thursday variable $X_3$ and the baseline Monday-Tuesday indicator, as both correspond to weekdays.

\begin{table}[H]
\centering
\caption{{$P$-values from the proposed global and partial tests.
The global test assesses the overall \F regression effect of the full
predictor vector $\bX$ on the transport network response $Y$.
The predictors are average trip distance ($X_1$), average fare per trip
($X_2$), and indicators for Wednesday--Thursday ($X_3$), Friday ($X_4$),
Saturday ($X_5$), and Sunday ($X_6$), with Monday--Tuesday as the
baseline category. The partial test for $X_j$ assesses its additional
effect conditional on the remaining predictors $\bX\setminus X_j$.}}
\label{tab:Taxi_pvalues}
\begin{tabular}{lccccccc}
\hline
& Global & \multicolumn{6}{c}{Partial} \\
\cline{2-2}\cline{3-8}
Predictor & $\mathbf{X}$ & $X_1$ & $X_2$ & $X_3$ & $X_4$ & $X_5$ & $X_6$ \\
\hline
$p$-value & $<0.001$ & $0.014$ & $0.212$ & $0.095$
          & $<0.001$ & $<0.001$ & $0.006$ \\
\hline
\end{tabular}
\end{table}



\begin{figure}[h!]
    \centering
    \includegraphics[width=\linewidth]{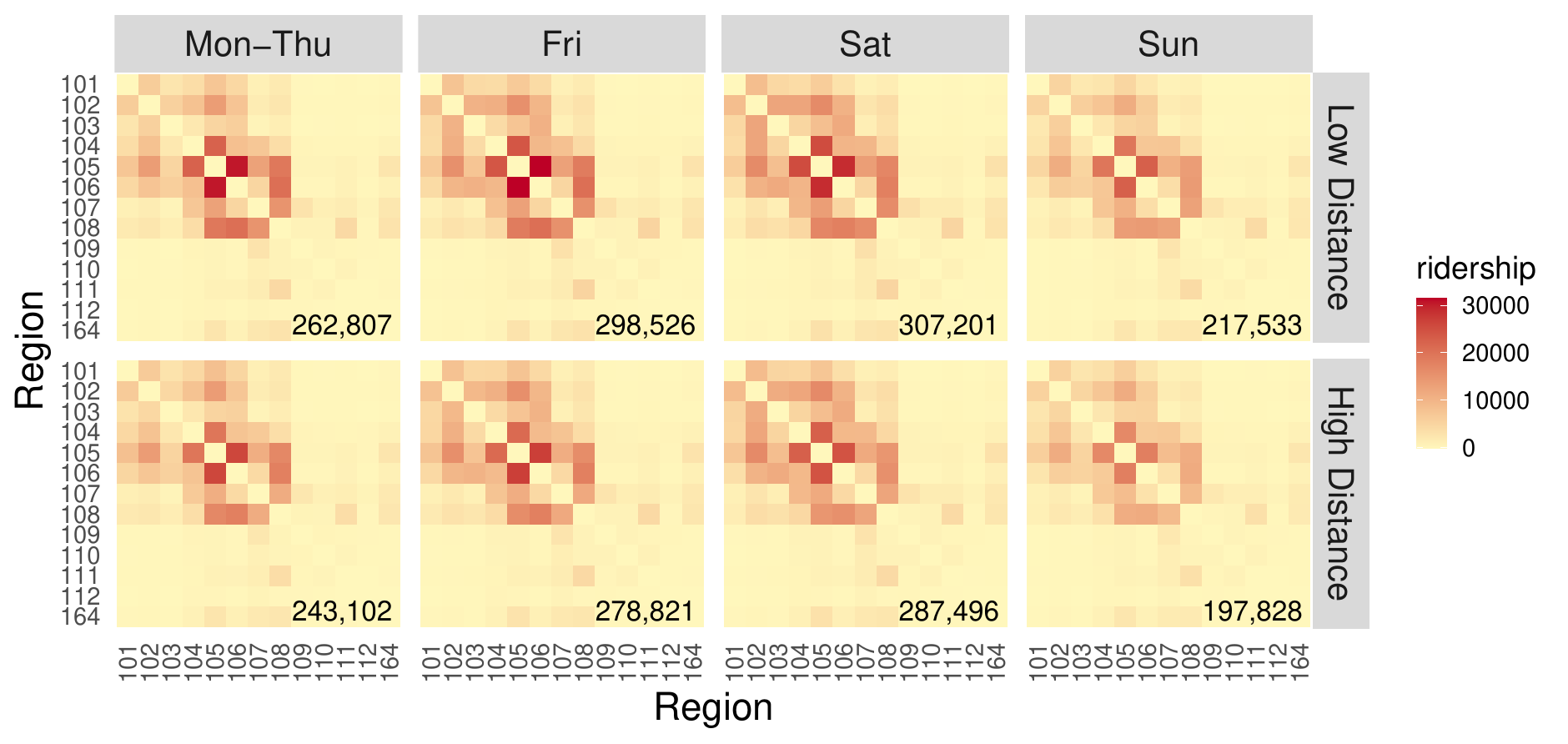}
    \caption{Estimated transport networks from the global network regression with reduced four-dimensional predictors (Friday, Saturday, Sunday, and average trip distance). Columns correspond to predicted networks for Monday-Thursday, Friday, Saturday, and Sunday, respectively, while rows represent networks at the 10th and 90th percentiles of trip distance. The value shown in the bottom-right corner of each panel indicates the total ridership across the entire transport network. }
    \label{fig:pred_networks}
\end{figure}

Motivated by these observations, we exclude $X_2$ (fare) and  $X_3$ (Wednesday-Thursday) and proceed with a reduced four-dimensional predictor vector: $X_1$ (average trip distance), $X_4$ (Friday), $X_5$ (Saturday), and $X_6$ (Sunday). We then fit the global \F regression to the transport network response $Y$ using this reduced set of predictors, under the same settings as in our simulation studies. Figure~\ref{fig:pred_networks} displays the estimated transport networks as heatmaps under different predictor evaluations. The columns correspond to predicted networks for Monday-Thursday, Friday, Saturday, and Sunday, respectively, while the two rows represent networks at the 10th and 90th percentiles of trip distance. The value shown in the bottom-right corner of each panel indicates the total ridership across the entire transport network. The predicted networks reveal clear differences across both day of week and trip distance levels. In particular, networks evaluated at 90th percentile of trip distances generally exhibit lower overall ridership compared to those evaluated at the 10th percentile of distances, with noticeably reduced connectivity between adjacent regions. In terms of weekday effects, total ridership is highest on Saturday and lowest on Sunday.


\subsection{U.S. energy source compositional data}\label{sec:energy}

Compositional data are ubiquitous and are not situated in a vector space as they correspond to vectors with non-negative elements that sum to $1$. Here, we focus on energy sources data for electricity generation across the United States, expressed as fractions or percentages, which are available at \url{https://www.eia.gov/electricity/data/state/}. In our analysis, we consider three categories of U.S. electricity generation energy sources from each state in one year: 1) $U_1$: Proportion of fossil fuels (coal or petroleum), 2) $U_2$: Proportion of natural gas, and 3) $U_3$: Proportion of all other sources, including nuclear, geothermal, conventional hydroelectric, and solar thermal.

Given the compositional nature of the data, we represent each data point as a point on the unit sphere $\mathbb{S}^2$ by applying a square root transformation described in Example 2 of Section \ref{sec:prelim}. Specifically, the annual energy-source composition of each state is represented on $\mathbb{S}^{2}$ as $Y = ({U_1^{1/2}}, {U_2^{1/2}},{U_3^{1/2}}),$ for the 50 U.S. states (excluding the District of Columbia) over the 8 years from 2012 to 2019. This yields a total of $n = 400$ observations of $Y.$

We considered four predictors ($p = 4$): The predictor $X_1$ represents the year, ranging from 2012 to 2019, and is scaled as $X_1 = (\text{year} - 2012)/7$; $X_2$ is the median household income (in units of $100{,}000$ US dollars) for each state in each year, obtained from the U.S. Census Bureau's 1-Year American Community Survey (ACS), accessible at \url{https://data.census.gov/table}; $X_3$ is a political vote margin defined as the difference in vote shares between the Republican and Democratic parties in U.S. House of Representatives general elections. For even-numbered years, we use actual state-level election results from
\url{https://dataverse.harvard.edu/dataverse/medsl_house}, and for odd-numbered years the  margin is interpolated from the two nearest election years; $X_4$ is an indicator of whether coal was mined in the state in that year, available at \url{https://www.eia.gov/coal/data}.

Following the previous section, we conduct a global test to assess whether the predictor vector $\bX = (X_1, \ldots, X_4)$ has \F regression effect on the U.S. electricity compositional response $Y$. To further investigate variable specific contributions, we perform partial \F regression tests. For each predictor $X_j$, we partition the predictors into $\bX_1 = \bX \setminus X_j$ and $\bX_2 = X_j$, and test whether inclusion of $X_j$ provides additional explanatory power beyond $\bX_1$. For test statistics $C_G$ in \eqref{eq:test_stat_Cauchy} and $C_P$ in \eqref{eq:test_stat_Cauchy_partial}, we use $K = 100$ Cauchy combination components with equal weights $c_k = 1/K,$ as in Section \ref{sec:simul}.

Table~\ref{tab:Energy_pvalues} shows the resulting $p$-values. The global test rejects the global null hypothesis of no \F regression effect at $\alpha = 0.05.$ The partial tests  indicate that the state-level coal mining variable provides substantial additional explanatory power beyond the other predictors, whereas vote margin, year, and household income do not show significant additional contributions when conditioning on the remaining predictors.

\begin{table}[htbp]
\centering
\caption{ {$P$-values from the proposed global and partial tests.
The global test assesses the overall \F regression effect of the full
predictor vector $\bX$ on the U.S. energy source compositional response $Y$.
The predictors are calendar year ($X_1$), median household income ($X_2$), political vote margin ($X_3$), and indicator of whether coal was mined in the state ($X_4$). The partial test for $X_j$ assesses its additional
effect conditional on the remaining predictors $\bX\setminus X_j.$}}
\label{tab:Energy_pvalues}
\begin{tabular}{lccccc}
\hline
& Global & \multicolumn{4}{c}{Partial} \\
\cline{2-2}\cline{3-6}
Predictor & $\mathbf{X}$ & $X_1$ & $X_2$ & $X_3$ & $X_4$ \\
\hline
$p$-value & $<0.001$ & $0.635$ & $0.332$ & $0.146$ & $0.003$ \\
\hline
\end{tabular}
\end{table}

\begin{figure}[h!]
    \centering
\includegraphics[width=0.45\linewidth]{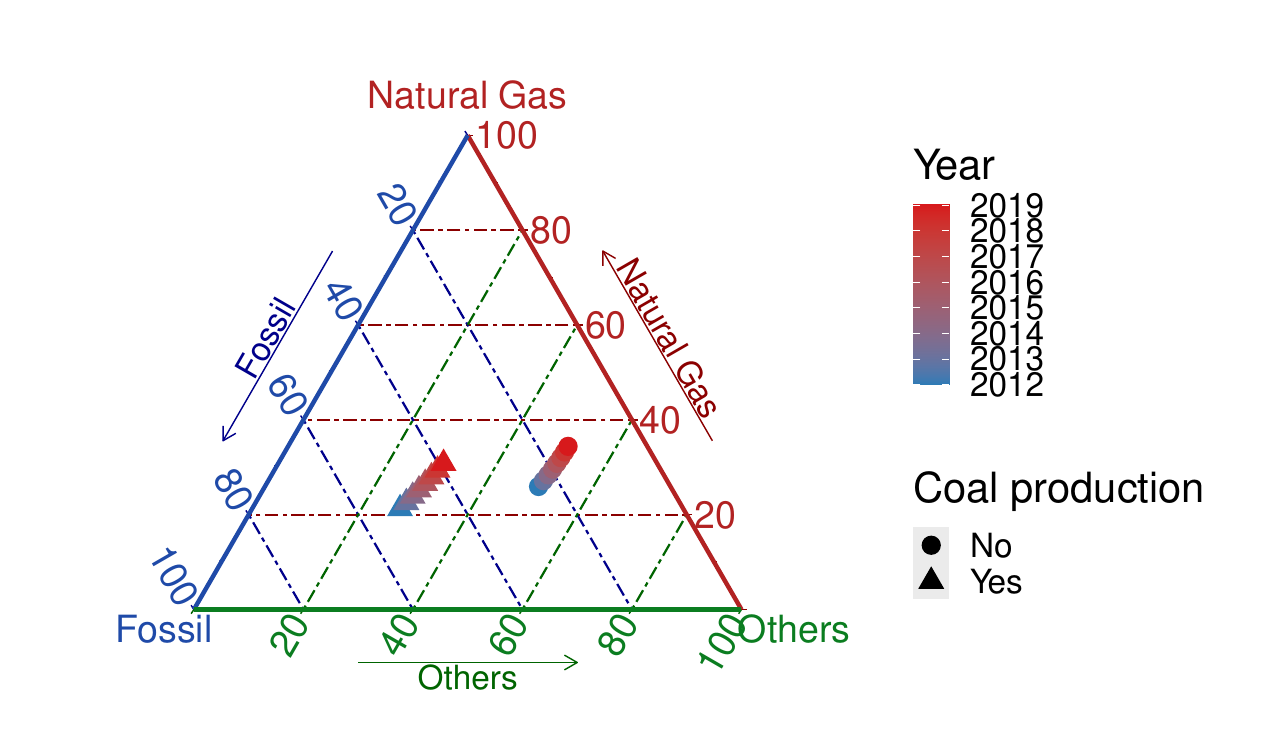}
\includegraphics[width=0.45\linewidth]{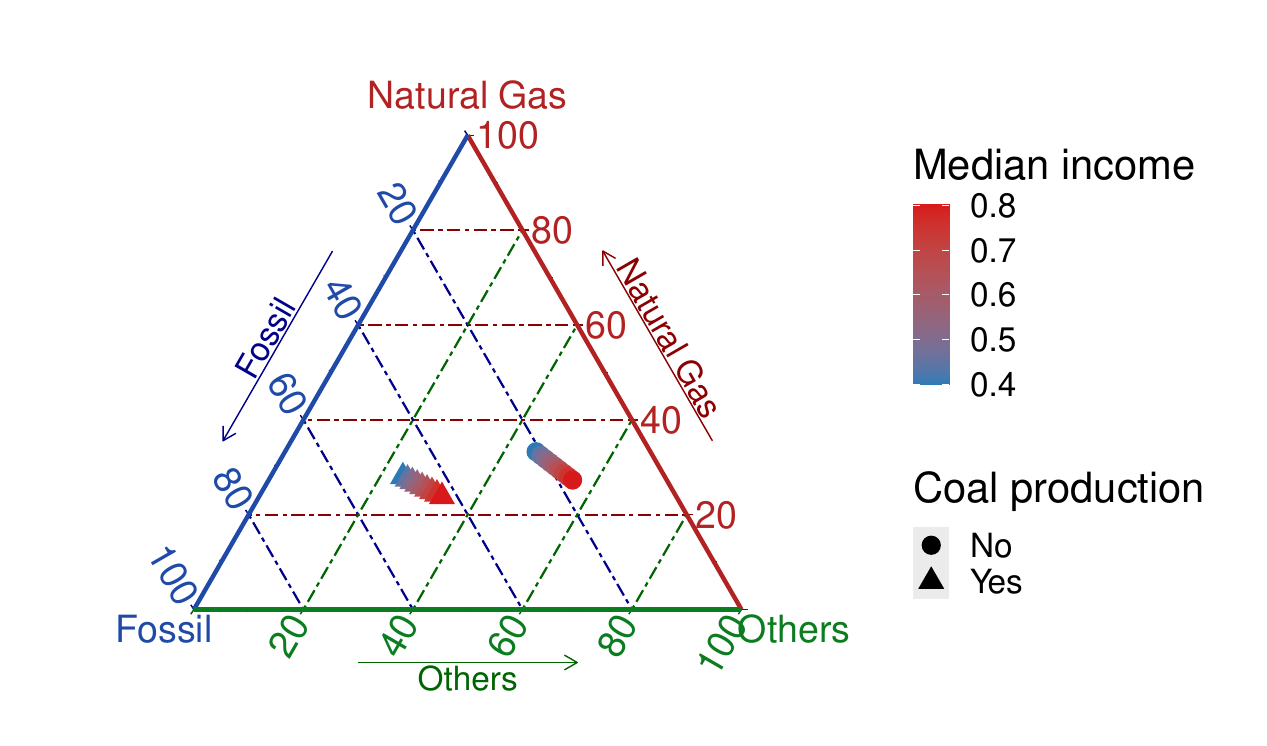}
\includegraphics[width=0.45\linewidth]{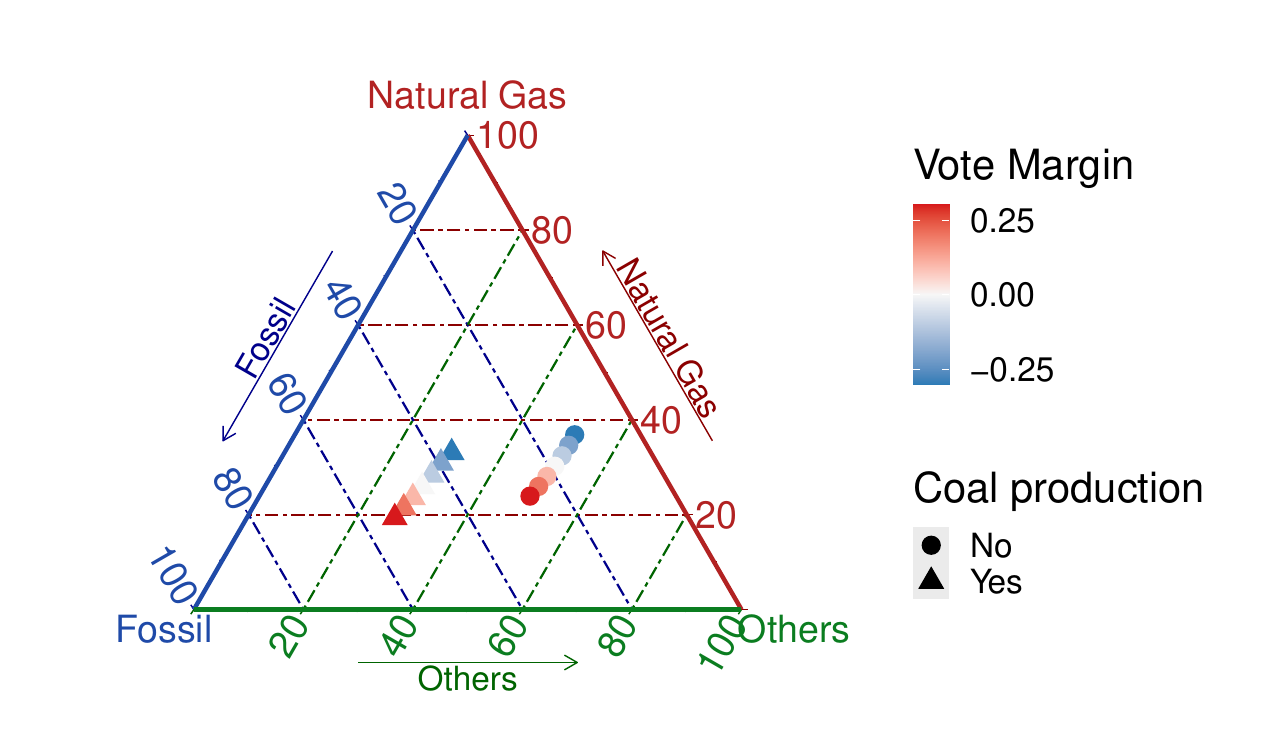}
    \caption{Estimated U.S. electricity generation compositions from global \F regression with four predictors $\bX$, visualized using ternary plots: (Top left) Estimated compositions as functions of year $X_1$ and coal production status $X_4$, with median household income fixed at $X_2 = 0.6$ (in units of \$100{,}000) and vote margin set to $X_3 = 0$; (Top right) Estimated compositions as functions of median household income $X_2$ and coal production status $X_4$, with year fixed at $X_1 = 0.5$ and vote margin $X_3 = 0$; (Bottom) Estimated compositions as functions of vote margin $X_3$ and coal production status $X_4,$ with year fixed at $X_1 = 0.5$ and median household income fixed at $X_2 = 0.6.$ }
    \label{fig:ternary}
\end{figure}

Following the partial test results, Figure~\ref{fig:ternary} presents the estimated U.S. electricity generation compositions from global \F regression with four predictors $\bX$, visualized using ternary plots. The first panel shows the estimated compositions as functions of year $X_1$ and coal production status $X_4$, with median household income fixed at $X_2 = 0.6$ (in units of \$100{,}000) and vote margin set to neutral, i.e., $X_3 = 0$. The second panel shows the estimated compositions as a function of median household income $X_2$ and coal production status $X_4$, with scaled year $X_1 = 0.5,$ corresponding to year $2015,$ and vote margin $X_3 = 0$. The third panel presents the estimated compositions as functions of vote margin $X_3$ and coal production status $X_4$, with fixed $X_1 = 0.5$ and $X_2 = 0.6.$

Consistent with the partial test results, the coal-mining variable $X_4$ provides significant additional explanatory power. In all three ternary plots, states with $X_4 = 1$ (i.e., coal-producing states) show generally higher proportions of fossil fuels and lower proportions of alternative energy sources. Specifically, in the top-left panel, the proportion of fossil fuels decreases over time, accompanied by an increase in natural gas usage. The top-right panel shows that states with higher median household income tend to have a larger share of alternative energy sources. In the bottom panel, states with stronger Democratic preference (i.e., more negative vote margin) show lower proportions of fossil fuels and higher proportions of natural gas.

\section{Discussion}

This paper develops global and partial significance tests in \F regression with responses in general metric spaces. We introduce the \F $R$-squared as a measure of goodness-of-fit and incorporate it into the proposed hypothesis testing procedures. To the best of our knowledge, this provides the first unified approach for inference in \F regression without assuming a specific non-Euclidean space. The proposed methods are broadly applicable to a wide range of object-valued data, including networks and compositional data, and demonstrate strong finite-sample performance in simulations. Applications to NYC transport networks and U.S. energy composition further illustrate their practical utility in assessing the effects of predictors on these responses.

{ As discussed after Theorem~\ref{thm: power_global}, the broad applicability of the proposed tests to metric space-valued responses may entail a loss of local efficiency in specific metric spaces with stronger structure assumptions. For example, in an ordinary Euclidean linear regression model in \eqref{eq:linear_reg_Euclidean}, $D(P) = \bm{\beta}^\T_1 \Sigma \bm{\beta}_1,$ and consequently this condition corresponds to $\bm{\beta}_1$ converging to $0$ more slowly than
$n^{-1/4},$ under a nonsingular design covariance, whereas the classical F-test can detect coefficients of order $n^{-1/2}.$ Thus, the proposed test trades local efficiency for applicability to a broad class of response metric spaces $(\O, d)$.
Our objective is to develop an inference procedure applicable to a broad class of metric spaces $(\Omega,d)$. The proposed randomization provides a common asymptotic null calibration without requiring a linear, differentiable, or likelihood-based structure. This generality may come at the cost of local efficiency in particular spaces possessing additional structure. Developing more efficient significance tests under model- and geometry-specific assumptions, for example in smooth manifolds or Hadamard spaces, and comparing their local asymptotic power with that of the proposed test is an important direction for future research.}

\bco
\section*{Data Availability Statement}

The data that support the findings in Section \ref{sec:taxi} are openly available at \url{https://www.nyc.gov/site/tlc/about/tlc-trip-record-data.page}. The data that support the findings in Section \ref{sec:energy} are openly available at \url{https://www.eia.gov/electricity/data/state/},  \url{https://data.census.gov/table},  \url{https://www.eia.gov/coal/data}, and 
\newline \url{https://dataverse.harvard.edu/dataverse/medsl_house}.
\fi

\section*{Disclosure Statement}

The authors report there are no competing interests to declare.

\begin{singlespace}
\bibliography{paper-ref}
\end{singlespace}

\end{document}